\documentclass[twocolumn, prl, superscriptaddress,notitlepage]{revtex4-1}
\usepackage{graphicx}
\usepackage{physics}
\usepackage{epsfig}
\usepackage{color}
\usepackage{xspace}
\usepackage{array}
\usepackage{amsmath}
\usepackage{amsfonts}
\usepackage{ulem}
\usepackage[dvipsnames]{xcolor}
\usepackage[breaklinks,colorlinks,linkcolor=blue,citecolor=blue,urlcolor=blue]{hyperref}
\newcommand{\Poincare}{Poincar\'e\xspace}
\newcommand{\Schrodinger}{Schr\"odinger\xspace}
\newcommand{\di}{\textrm{d}}

\newcommand{\Li}{\textrm{Li}}

\begin{document}
\title{Curving the space by non-Hermiticity}
\author{Chenwei Lv}
\thanks{They contribute equally to this work.}
\affiliation{Department of Physics and Astronomy, Purdue University, West Lafayette, IN, 47907, USA}
\author{Ren Zhang}
\thanks{They contribute equally to this work.}
\affiliation{School of Physics, Xi'an Jiaotong University, Xi'an, Shaanxi 710049, China }
\affiliation{Department of Physics and Astronomy, Purdue University, West Lafayette, IN, 47907, USA}
\author{Zhengzheng Zhai}
\affiliation{Department of Physics and Astronomy, Purdue University, West Lafayette, IN, 47907, USA}
\author{Qi Zhou}
\email{zhou753@purdue.edu}
\affiliation{Department of Physics and Astronomy, Purdue University, West Lafayette, IN, 47907, USA}
\affiliation{Purdue Quantum Science and Engineering Institute, Purdue University, West Lafayette, IN, 47907, USA}
\date{\today}
\begin{abstract}

Quantum systems are often classified into Hermitian and non-Hermitian ones. Extraordinary non-Hermitian phenomena, ranging from the non-Hermitian skin effect to the supersensitivity to boundary conditions, have been widely explored. Whereas these intriguing phenomena have been considered peculiar to non-Hermitian systems, we show that they can be naturally explained by a duality between non-Hermitian models in flat spaces and their counterparts, which could be Hermitian, in curved spaces. For instance, prototypical one-dimensional (1D) chains with uniform chiral tunnelings are equivalent to their duals in two-dimensional (2D) hyperbolic spaces with or without magnetic fields, and non-uniform tunnelings could further tailor local curvatures. Such a duality unfolds deep geometric roots of non-Hermitian phenomena, delivers an unprecedented routine connecting Hermitian and non-Hermitian physics, and gives rise to a theoretical perspective reformulating our understandings of curvatures and distance.  In practice, it provides experimentalists with a powerful two-fold application, using non-Hermiticity as a new protocol to engineer curvatures or implementing synthetic curved spaces to explore non-Hermitian quantum physics. 

\end{abstract}

\maketitle

System-environment couplings lead to a plethora of intriguing non-Hermitian phenomena~\cite{Syassen2008,Regensburger2012,Feng2014,Longhi2015,Chen2017,LuoL2019,Yan2020,Weidemann2020,Helbig2020}, such as non-orthogonal eigenstates, the non-Hermitian skin effect~\cite{Wang2018,Kunst2018,Torres2018,Thomale2019,Slager2020,Fang2020,Sato2020}, real energy spectra in certain parameter regimes~\cite{Bender1998,Mostafazadeh2002}, collapsed energy spectra and coalesced eigenstates at an exceptional point (EP)~\cite{Torres2018,Ashida2020,Bergholtz2021}, and drastic responses to boundary conditions~\cite{Xiong2018,Sato2019}.
While these phenomena have been extensively explored in quantum sciences and technologies~\cite{Regensburger2012,Wiersig2014,Feng2014,Longhi2015,Chen2017,Kawabata2019,LeeJY2019,LeeJY2019-2,Weidemann2020,Xiao2020}, peculiar theoretical tools are often required to study non-Hermitian physics.  
Though bi-orthogonal vectors and metric operators are introduced to restore orthogonality~\cite{Brody2013,Kunst2018,Ashida2020, Scholtz1992,Mostafazadeh2010}, the underlying physics of these mathematical tools is not clear yet.
Moreover, it remains challenging to prove the real energy spectra of certain non-Hermitian systems, as the existence of the $\mathcal{PT}$ symmetry does not guarantee a real energy spectrum and sophisticated mathematical techniques are required~\cite{Bender1998, Mostafazadeh2002, Dorey2004}.  

Here, we point out a duality between non-Hermitian Hamiltonians in flat spaces and their counterparts in curved spaces. 
On the theoretical side, this duality leads to a geometric framework providing a unified explanation of several
non-Hermitian phenomena. 
For instance, it is the finite curvature that requires an orthonormal condition distinct from that in flat spaces, enforces eigenstates to localize at edges, and gives rise to the supersensitivity to boundary conditions. 
Dual models in curved spaces could be Hermitian, providing a simple proof of the existence of real energy spectra in certain non-Hermitian systems. 
Moreover, in sharp contrast to existing schemes of studying curved spaces~\cite{Bekenstein2017,ZhouZW2018,Schine2019,Kollr2019}, which were built on the conventional wisdom that a flat space needs to be physically distorted to become curved, our results show that non-Hermiticity is a controllable knob for tuning curvatures  even when the space appears to be flat, for instance, in lattices with fixed lattice spacing. This duality therefore may reform our understandings of distance and curvatures. 
 
In practice, our duality has a two-fold implication. On the one hand, it establishes non-Hermiticity as a unique tool to simulate intriguing quantum systems in curved spaces.  
For instance, it offers a new approach of using non-Hermitian systems in flat spaces to solve the grand challenge of accessing gravitational responses of quantum Hall states (QHS) in curved spaces~\cite{Wen1992,Zograf1995,Wiegmann2014,Schine2019}. 
On the other hand, the duality allows experimentalists to use curved spaces to explore non-Hermitian physics. 
Whereas a variety of non-Hermitian phenomena have been addressed in experiments, delicate designs of dissipations are often required~\cite{Nelson1993,Regensburger2012,Longhi2015,Nelson2016,Zoller2017,Yan2020,Weidemann2020,Helbig2020,Yang2021}. 
Our results show that curved spaces can serve as an unprecedented means to explore non-Hermitian Hamiltonians without resorting to dissipations.

\begin{figure}[t] 
  \includegraphics [angle=0,width=.5\textwidth]
  {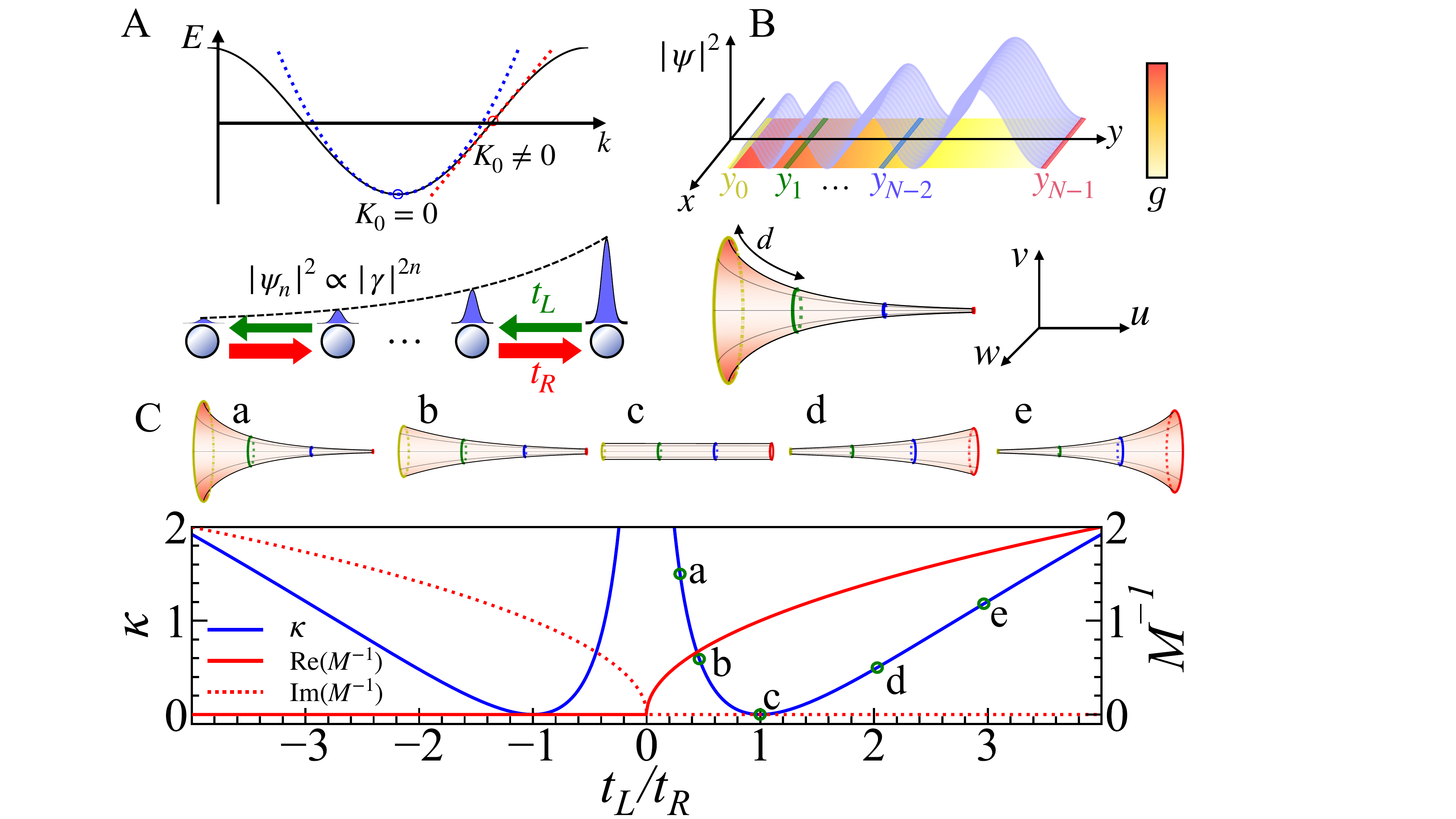}
  \caption {The duality between the Hatano-Nelson (HN)
  model and a hyperbolic surface. 
  (A) A HN chain and its energy spectrum as a function of $k$. 
  Near a vanishing (finite) $K_0$, the effective theory in curved space is non-relativistic (relativistic).
  Eigenstates on the HN chain
  are localized at the edge, $|\psi_n|^2\propto |\gamma|^{2n}$.
  (B)A HN chain is mapped to the shaded strip on the \Poincare half-plane,
  in which an eigenstate with $k_x=0$  satisfies $|\psi|^2\propto y$. 
  This shaded strip on the \Poincare half-plane with PBC in the $x$-direction
  is equivalent to a pseudosphere embedded in 3D Euclidean space. 
  (C)   The curvature and the inverse of the effective mass, as functions of $t_L$ for a fixed $t_R$.
   The unites of $\kappa$ and $M^{-1}$  are $1/d^2$ and $2t_Rd^2/(\hbar^2)$, respectively.
  (a-e) show the dual pseudospheres of the HN model at various $t_L>0$. A pseudosphere for $t_L<0$ is the same as that for $-t_L$.
  }\label{fig:fig1}
\end{figure}

Our duality can be demonstrated using the celebrated 
Hatano-Nelson (HN) model~\cite{Hatano1996}, which reads, 
\begin{equation}
-t_R\psi_{n-1}-t_L\psi_{n+1}=E\psi_n, \label{HN}
\end{equation}
where $n=0,1,...N-1$ is the lattice index of a one-dimensional (1D) chain, $\psi_n$ is the eigenstate, $E$ is the corresponding eigenenergy, and $t_L$ and $t_R$ are the nearest-neighbor tunneling amplitudes towards the left and the right, respectively. 
Under the open boundary condition (OBC), $\psi_n=e^{{n}\ln(\gamma)}\sin(k_mnd)/\sqrt{(N-1)/2}$, where $\gamma=\sqrt{{t_R}/{t_L}}$ characterizes the strength of non-Hermiticity, $k_m=m\pi/((N-1)d)$, $m\in\mathbb{Z}$, and $d$ is the lattice constant.
The eigenenergy reads $E_m=-2\sqrt{t_Lt_R}\cos(k_md)$. 

Similar to Hermitian lattice models, the effective theory of  Eq.~(\ref{HN}) in the continuum limit describes the motion of a non-relativistic (relativistic) particle at (away from) the band bottom and top, with a quartic (linear) dispersion relation, as shown in Fig. (1A). 
At the band bottom, the effective theory is written as 
\begin{equation}
   -\frac{\hbar^2}{2M}\kappa \left(y^2\partial_y^2+\frac1{4}\right)\psi(y)=E\psi(y),\label{1DPH}
\end{equation}
where $M=\hbar^2/(2\sqrt{t_Lt_R}d^2)$ is the effective mass, and $\kappa=4 \ln^2(|\gamma|)/d^2$. 
Solutions to Eq.(\ref{1DPH}),  $y^{\frac{1}{2}}y^{\pm ik_y/\sqrt{\kappa}}$, have the same energy, $\hbar^2 k_y^2/(2M)$.
An eigenstate under OBC is their superposition, $\psi(y)=\sqrt{2/\ln(y_{N-1}/y_0)} (y/y_0)^{\frac{1}{2}}\sin[k_y\ln(y/y_0)/\sqrt{\kappa}]$ with $k_y=m\pi\sqrt{\kappa}/\ln(y_{N-1}/y_0)$ and $\psi(y_0)=\psi({y_{N-1}})=0$. $y_0$ and $y_{N-1}$ specify positions of the two edges. 
At the band top,  we have $M\rightarrow -M$.
Eq.(\ref{1DPH}) is a dimension reduction of the \Schrodinger equation on a Poincar\'e half-plane, 
\begin{equation}
-\frac{\hbar^2}{2M}\kappa \left(y^2\nabla^2+\frac1{4}\right)\Psi(x,y)=E\Psi(x,y)\label{PH},
\end{equation}
where $\nabla^2\equiv\left(\partial_x^2+\partial_y^2\right)$, $\Psi(x, y)=e^{ik_x x} \psi (y)$, and $-\kappa$ is the curvature (Supplementary Material). 
The metric tensor is ${\bf g}=\frac{1}{\kappa y^2}(\di x^2+\di y^2)$, $g=\det({\bf g})=1/(\kappa^2 y^4)$.
Since $k_x$ is a good quantum number, Eq.(\ref{PH}) reduces to Eq. (\ref{1DPH}) when $k_x=0$. 
A finite $k_x$ adds an onsite potential to the HN model,
\begin{equation}
   V_n\psi_{n}-t_R\psi_{n-1}-t_L\psi_{n+1}=E\psi_n,\label{LM}
\end{equation}
where $V_n=a^2\sqrt{t_Rt_L}\gamma^{4n}$. The dimensionless quantity $a^2=4 (\ln^2|\gamma|) y_0^2k_x^2$ characterizes the strength of $V_n$.

\begin{table*}[t]
  \includegraphics[width=1.\textwidth]{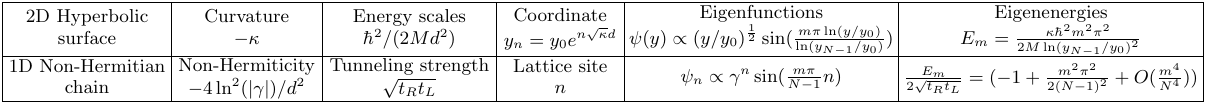}
\caption{{The mapping between the continuum limit of the HN model near the band bottom under OBC and the \Poincare half-plane. }
}\label{table1}
\end{table*}

To derive the duality between the continuum limit of Eq.~(\ref{HN}) at the band bottom and Eq.~(\ref{1DPH}), we define $\psi_n\equiv \sqrt{d}\psi(s_n)$ with $s_n=nd$, 
such that the eigenstate of the HN model, $\psi_n$, defined on discrete lattice sites is extended to $\psi(s)$ as a function of a continuous variable $s$. 
Since $\psi_n$, under OBC, includes a part that changes exponentially, i.e., $e^{n\ln(\gamma)}$, so does $\psi$.
We thus define $\phi(s)\equiv\psi(s)e^{-qs}$ with $q=\ln(\gamma)/d=\frac{1}{2d}\ln (t_R/t_L)$ determining the inverse of the localization length, and $\phi(s)$ varies slowly with changing $s$.  
Then we have $\psi_n=\sqrt{d}\phi(s_n)e^{qs_{n}}$.
Substituting $\psi_n$ into Eq.~(\ref{HN}) and using the Taylor expansion for $\phi(s)$, $\phi(s_{n\pm 1})=\phi(s_n)\pm d \partial_s \phi +\frac{1}{2}d^2 \partial^2_s \phi$, we obtain $-\sqrt{t_Lt_R}(2+\partial_s^2)\phi=E\phi$. 
Consequently, $\psi(s)$ satisfies
\begin{equation}
    -\sqrt{t_Lt_R}d^2\left(\partial^2_s-2 q\partial_s+q^2+2/{d^2}\right)\psi(s)=E\psi(s) \label{HNc2}.
\end{equation}
It describes a nonrelativistic particle subject to an imaginary vector potential, $\vec{A}\sim i q$~\cite{Nelson1993,Hatano1996}. Unlike a real vector potential that amounts to a $U(1)$ gauge field, here, an imaginary vector potential curves the space. 
Performing a coordinate transformation $y/y_{0}=e^{2qs}$ and defining $M=\hbar^2/(2\sqrt{t_Lt_R}d^2)$, $\kappa=4 \ln^2(|\gamma|)/d^2$, we obtain Eq.~(\ref{1DPH}) up to a constant energy shift $-2\sqrt{t_Lt_R}$. 
The mapping between these two models is summarized in Table I, which provides a dictionary translating microscopic parameters between them.
For instance, $\psi_n$, the wavefunction at the $n$-th lattice site of the NH model is identical to $\psi(y_n)$, the wavefunction on the Poincar\'e half-plane evaluated at $y_n=y_0e^{n\sqrt{\kappa}d}$. 
The low-energy limit of the eigenenergy of Eq.~(\ref{HN}) is also identical to the eigenenergy of Eq.~(\ref{1DPH}) as shown by Table I. 

Away from the band bottom(top), similar calculations can be performed by defining $\psi(s)=e^{\pm iK_0s}e^{qs}\phi(s)$ using Taylor expansions of the slowly varying $\phi(s)$ and the same coordinate transformation $y=y_0e^{2qs}$. 
We obtain the effective theory near $K_0d\neq 0,\pm\pi$,
\begin{equation}
 [ E(K_0) \pm i\sqrt{\kappa}\hbar v_Fy\left(\partial_y-1/(2y)\right)]\psi(y)=E\psi(y), \label{NR}
\end{equation}
where $E( K_0)=-2\sqrt{t_Lt_R}(\cos(K_0d)+K_0d\sin(K_0d))$, $v_F=-2\sqrt{t_Lt_R}d\sin(K_0d)/\hbar$, and $\pm$ corresponds to the left and right moving waves centered near $\pm K_0$, respectively, The previously defined $\kappa=4\ln^2(|\gamma|)/d^2$ has been used. 
 The eigenstate under OBC includes both the left and right moving waves and is written as $\sqrt{2/\ln(y_{N-1}/y_0)} (y/y_0)^{\frac{1}{2}}\sin[k_y\ln(y/y_0)/\sqrt{\kappa}]$ with eigenenergy of $-2\sqrt{t_Lt_R}[\cos(K_0d)+(K_0-k_y)d\sin(K_0d)]$, which recovers the results of the HN model near a finite $K_0$. 

Our duality provides a natural explanation of several peculiar non-Hermitian phenomena. 
Firstly, the orthonormal condition of effective theories in Eq.(\ref{1DPH}) and Eq.(\ref{NR}) reads
\begin{equation}
    \int\frac{\di y}{\kappa y^2}\psi^*(y; k_y)\psi(y; k^\prime_y)=\mathcal{N}\delta_{k_y,k^\prime_y},\label{NC}
\end{equation}
where the normalization constant $\mathcal{N}$ can be chosen freely.
As a common feature of curved spaces, a finite curvature appears in the above equation. 
Considering a strip in the domain $x_0\le x\le x_0+L$, its width in the $x$-direction depends on $y$, $L_x(y)=\int_{x=x_0}^{x_0+L}\di x/(\sqrt{\kappa}y)=L/(\sqrt{\kappa}y)$.
Thus, a wave packet traveling in the $y$-direction must include an extra factor $y^{\frac{1}{2}}$ to guarantee the conservation of particle numbers. 
In the $s$-coordinate, Eq.~(\ref{NC}) is written as $\int\frac{\di s}{\sqrt{\kappa}y_0}e^{-2qs}\psi_{k_{y}}^*(s)\psi_{k^\prime_{y}}(s)=\mathcal{N}\delta_{k_{y},k^\prime_{y}}$.
Discretizing this equation with $\mathcal{N}=(\sqrt{\kappa} y_0)^{-1}$, and transforming it to the HN model, we obtain,
\begin{equation}
    \sum_{n}|\gamma|^{-2n}\psi^*_n(k_m)\psi_n(k_{m^\prime})=\delta_{k_m,k_{m^\prime}}, \label{NCL}
\end{equation}
where $|\gamma|^{-2n}$
is precisely the difference between the
left and right eigenvectors, or the metric operator~\cite{Mostafazadeh2010}. 
The mapping to a curved space thus establishes an explicit physical interpretation of orthonormal conditions in non-Hermitian systems. 

Secondly, the duality allows us to equate the non-Hermitian skin effect to its counterpart on the \Poincare half-plane we found recently~\cite{Zhou2020}. 
This can be best visualized using the embedding of  a hyperbolic surface in three-dimensional (3D) Euclidean space. 
We define $y=r_0\cosh(\eta)$, $x=r_0\varphi$, where $r_0$ is an arbitrary constant and $\eta>0$, $\varphi\in(-\pi,\pi)$. 
The embedding can then be written as
\begin{equation}
   (u,v,w)=\frac{1}{\sqrt{\kappa}}\bigg((\eta-\tanh(\eta)),\frac{\cos(\varphi)}{\cosh(\eta)},\frac{\sin(\varphi)}{\cosh(\eta)}\bigg).
     \label{ps}
\end{equation}
This is a parameterization of a pseudosphere with a constant negative curvature and a radius of $1/\sqrt{\kappa}$, which satisfies $(u-{\rm arcsech}(\sqrt{(v^2+w^2)\kappa})/\sqrt{\kappa})^2+v^2+w^2=\kappa^{-1}$. 
As shown in Fig.~\ref{fig:fig1} B, a pseudosphere features a funnel shape, since the circumference of the circle with a fixed $y(\eta)$ changes with changing $y(\eta)$.
As previously explained, a coordinate transformation $y=y_0e^{2qs}$ maps eigenstates on the hyperbolic surface, $y^{\frac{1}{2}}y^{ik_{y}/\sqrt{\kappa}}$, to $e^{qs}e^{i(2q/\sqrt{\kappa})k_ys}$, which exponentially localizes near the funneling mouth, the smaller end. 

Thirdly, the collapsed energy spectrum at EP of the HN models has a natural geometric interpretation. 
When $t_L=t_R$, the pseudosphere reduces to a cylinder with a vanishing $\kappa$. 
For a given $t_{R}$($>t_{L}$), $\kappa$ increases with decreasing $t_L$. 
Increasing the non-Hermiticity thus makes the space more curved, as shown by Fig.~\ref{fig:fig1} C.
Approaching EP, $t_L\rightarrow 0$, $\kappa$ diverges, and the localization length, $1/\ln(|\gamma|)$, vanishes, forcing all eigenstates to coalesce.  
As eigenenergies read $E=\hbar^2 k_y^2/(2M)$ with  divergent $M$, eigenenergies collapse to zero with a massive degeneracy. 
Across EP, $t_Lt_R<0$, and the effective mass becomes imaginary, all previous results of positive $t_Lt_R$ still apply provided that $M\rightarrow \pm i M$. 
Particles moving in hyperbolic spaces are thus dissipative, and stationary states no longer exist.

Lastly, similar to the HN model, changing OBC to PBC leads to drastic changes in the curved space.
Eigenstates of Eq.(\ref{1DPH}) and Eq.(\ref{NR}) normalized to $\mathcal{N}$ become $\sqrt{\kappa \mathcal{N}}(y_0^{-1}-y_{N-1}^{-1})^{-\frac1{2}}(y/y_0)^{ik_{y}/\sqrt{\kappa}}$, where $k_y=2m\pi\sqrt{\kappa}/\ln(y_{N-1}/y_0))$ so that $\psi(y_0)=\psi(y_{N-1})$.
Correspondingly, eigenenergies become complex. 
This can be explicitly shown from the time-dependent \Schrodinger equations. 
For instance, at the band bottom(top), we multiply $\psi^*(y)$ to both sides of $i\hbar\partial_t\psi=-\frac{\hbar^2\kappa }{2M}\left(y^2\partial_y^2+1/4\right)\psi$, subtract from the resultant expression its complex conjugate, and integrate over $y$ from $y_0$ to $y_{N-1}$.
We find that the total particle number $ \mathcal{N}_{p}=\int_{y_0}^{y_{N-1}}\di y|\psi(y)|^2/(\kappa y^2)$ satisfies, 
\begin{equation}
  \partial_t \mathcal{N}_{p}={\hbar \sqrt{\kappa}\mathcal{N}k_y}/M,\label{PN}
\end{equation}
which signifies the absence of a stationary state and explains complex eigenenergies under PBC.
Using $\partial_t \mathcal{N}_p=\frac{2}{\hbar}{\rm Im}(E)\mathcal{N}_p$, we find ${\rm Im}(E)=\hbar^2\sqrt{\kappa}k_y/(2M)$.
This is distinct from the result for OBC, where $\psi(y)\sim y^{1/2}y^{\pm ik_y/\sqrt{\kappa}}$ such that $\partial_t \mathcal{N}_{p}=0$. 
Similar calculations can be performed for effective theories away from the band top (bottom), $i\hbar\partial_t\psi=\left[E(K_0) + i\sqrt{\kappa}\hbar v_Fy\left(\partial_y-1/(2y)\right)\right]\psi$. 
Straightforward calculations show that $\partial_t\mathcal{N}_p=-\sqrt{\kappa}v_F\mathcal{N}$, which explains the imaginary part of the eigenenergy, ${\rm Im}(E)=-\sqrt{\kappa} \hbar v_F/2$. 

\begin{figure}[t] 
    \includegraphics [angle=0,width=0.5\textwidth]
    {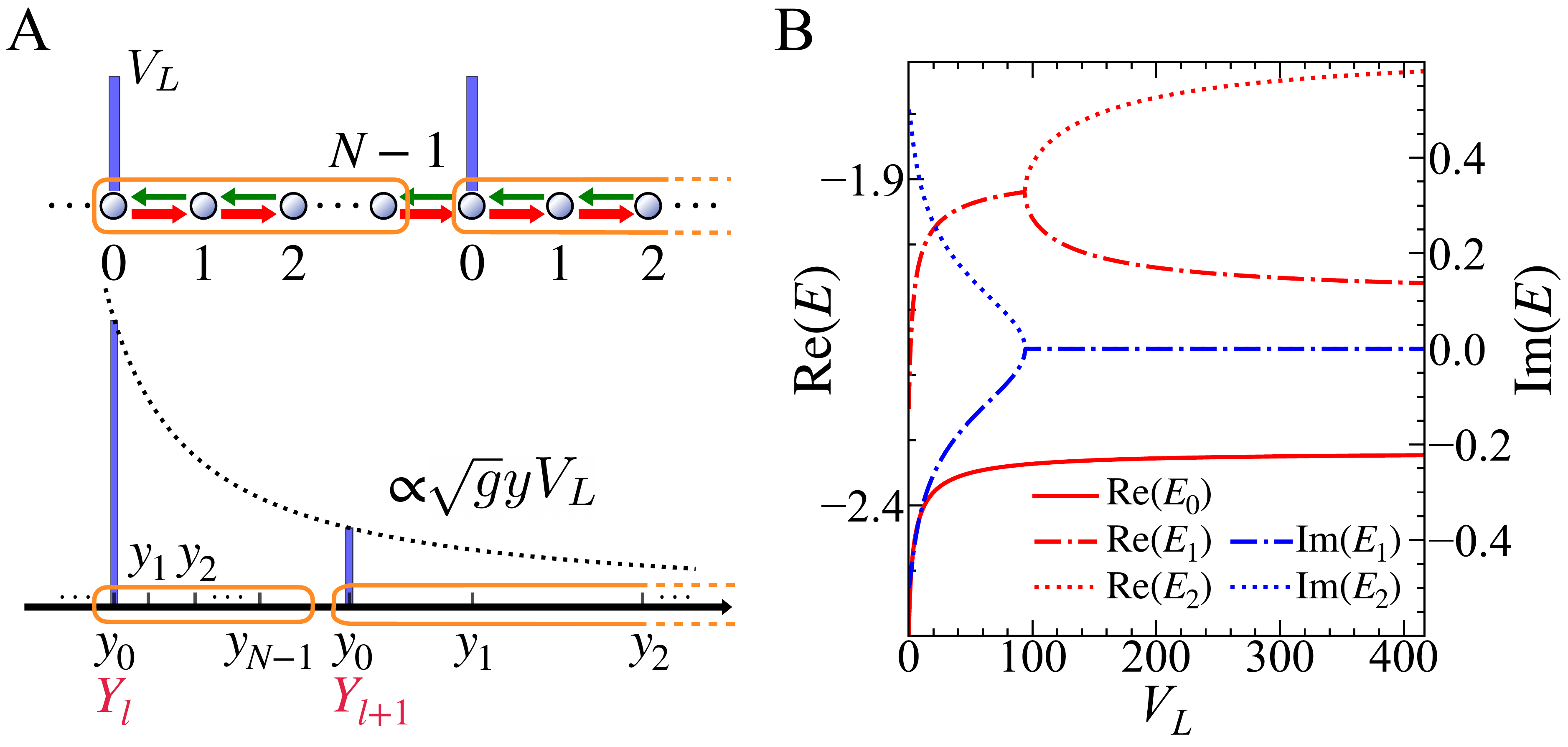}
    \caption
    {Changing boundary conditions. (A) 
    A constant potential, $V_L$, is added to the first site in each unit cell of the superlattice. 
    The corresponding potential in the curved space depends on the position. 
    (B) Eigenenergies for the ground state $E_0\in\mathbb{R}$ and the first two excited states $E_{1,2}\in\mathbb{C}$ as functions of $V_L$.
    $\sqrt{t_R/t_L}=1.5$ and $N=12$ are used. }\label{fig:fig2}
  \end{figure}

The boundary condition can be continuously tuned. 
An onsite energy offset, $V_L\ge 0$, in one of the lattice sites of the HN model continuously changes PBC to OBC once $V_L$ increase from $0$ to $\infty$.
We consider a superlattice of a lattice spacing of $Nd$, whose unit cell is a HN chain, as shown in Fig.~\ref{fig:fig2} A. 
Fig.~\ref{fig:fig2}~B shows  eigenenergies as functions of $V_L$. 
Similarly,  an external potential can be added to the \Poincare half-plane,
\begin{equation}
  V_{\delta}=dV_L\sqrt{\kappa}y\sum_{l}\delta(y-Y_l),
\end{equation}
where $Y_l=y_0e^{Nl\sqrt{\kappa}d}$ is the lattice site of the superlattice. 
The $y$-dependent amplitude of the delta-functions guarantees the scale invariance and the equivalence between each section between $Y_l$ and $Y_{l+1}$. 
With $V_L$ increasing from zero to infinity, eigenstates evolve from those under PBC to the ones under OBC.
$\int_{Y_l^-}^{Y_l^+}\sqrt{g}\di yV_{\delta}\sim1/\sqrt{\kappa}$ sets the energy scale of the potential, such that the larger the non-Hermiticity is, the more sensitive of the system is to the boundary condition. 

Whereas the HN model provides an illuminating example of the duality, applications of our approach to generic non-Hermitian models are straightforward. We consider 
\begin{equation}
-\sum_{m-1}^M t_{Rm} \psi_{n-m}-\sum_{m=1}^M t_{Lm} \psi_{n+m}=E\psi_n,
\end{equation} 
where $t_{Rm}$ and $t_{Lm}$ are tunneling amplitudes from the $(n\mp m)$th to $n$th sites. 
An eigenstate under OBC in the bulk is written as $e^{ik nd+qnd}$, where $kd\in [0, 2\pi]$ and $q$ is real. 
Unlike the HN model, where $q=\ln(t_R/t_L)/(2d)$ is a constant, once beyond the nearest neighbor tunnelings exist, $q$ becomes a function of $k$ and defines the so-called generalized Brillouin zone (BZ) in the complex plane~\cite{Wang2018,Yokomizo2019,Hu2020,Fang2020,Xiao2020}. 
Near any point in the generalized BZ specified by $K_0d\in [0,2\pi]$, we define $\psi(s)=e^{iK_0s}e^{q(K_0)s}\phi(s)$, where $\phi(s)$ changes slowly as a function of $s$, corresponding to small deviations of the momentum in the continuum limit. 
Similar to discussions about the HN model, the effective theory can be formulated straightforwardly using $\phi(s_{n\pm 1})=\phi(s_n)\pm d \partial_s \phi +\frac{1}{2}d^2 \partial^2_s \phi$.  
The Schr\"odinger equation satisfied by $\psi(s)$ is written as 
\begin{equation}
-\mathcal{B}(K_0)[\partial^2_s-2\mathcal{A}(K_0)\partial_s+\mathcal{C}(K_0) ]\psi(s)=E\psi(s),
\end{equation}
where $\mathcal{A}(K_0)$, $\mathcal{B}(K_0)$ and $\mathcal{C}(K_0)$ depend on $K_0$, as well as $t_{Rm}$ and $t_{Lm}$. 
When only the nearest neighbor tunnelings exist, the above equation recovers Eq.(\ref{HNc2}) at $K_0=0$ and $\mathcal{A}(K_0)$ becomes real and reduces to a constant imaginary vector potential $\sim\ln (t_R/t_L)/(2d)$ that we have discussed in the HN model. 
In the most generic case, $\mathcal{A}(K_0)$ provides a complex vector potential, whose real part curves the space. 
Using a coordinate transformation $y=y_0e ^{2 \mathcal{A}_R(K_0) s}$, where $\mathcal{A}_R(K_0)$ is the real part of $\mathcal{A}(K_0)$, a hyperbolic surface is thus obtained in the same manner as the HN model. 
The only difference is that $\kappa$ now is written as $\kappa=4\mathcal{A}_R^2$ and depends on $K_0$. 
Such $K_0$-dependent curvature provides a geometric interpretation for the generalized BZ. 
Explicit calculations for a model including the next-nearest-neighbor interaction are given in Supplementary Materials.

Whereas uniform chiral tunnelings lead to a hyperbolic surface with a constant curvature, we could also consider non-Hermitian models with non-uniform tunnelings,  
\begin{equation}
   -t_{R,n-1}\psi_{n-1}-t_{L,n}\psi_{n+1}=E\psi_n\label{dschr},
\end{equation}
which gives rise to inhomogeneous local curvatures. 
For slowly varying $t_{R,n}$ and $t_{L,n}$, we define $\bar{t}(s)$ and $\bar{\gamma}(s)$ such that $\bar{t}(nd)=\frac{2Md^2}{\hbar^2}\sqrt{t_{R,n}t_{L,n}}$ and $\bar{\gamma}(nd)=\sqrt{t_{R,n}/t_{L,n}}$. 
We introduce a slowly changing function $\phi(s)=e^{\nu(s)/2}\psi(s)$ with $\nu(s)=\frac2{d}\int_0^{s}\ln(\bar{\gamma}(s'))\di s'$.  
This is a generalization of the uniform case, where $\nu(nd)$ reduces to a linear function of $n$, i.e., the previously discussed $n\ln (\gamma) $ in the HN model. 
Using the same procedure, we obtain the effective theory of Eq.(\ref{dschr}). 
For instance, the non-relativistic theory is written as 
\begin{equation}
   \frac{\hbar^2}{2M}\left(-\frac{1}{\sqrt{g}}\partial_ig^{ij}\sqrt{g}\partial_j-\frac{\kappa}{4}+V_c\right)\Psi(x,y)=E\Psi(x,y),\label{cshrdual}
\end{equation}
where $g_{xx}=g_{yy}=\sqrt{g}=\bar{t}(s_y)e^{-\frac4{d}\int_0^{s_y}\ln(\bar{\gamma}(s'))\di s'}$, $g_{xy}=g_{yx}=0$, 
$V_c=\frac{\hbar^2}{2Md^2}\left(\frac{d}{2}\partial_s\ln\bar{\gamma}|_{s_y}-2\right)\bar{t}(s_y)$, and the position-dependent curvature is written as $\kappa(y)=\bar{t}(s_y)\left(4\ln\bar{\gamma}^2(s_y)-2d\partial_s\ln\bar{\gamma}|_{s_y}\right)/d^2$. 
In these expressions, $s_y$ is obtained from $y-y_0=\int^{s_y}_0 \di s' e^{\frac2{d}\int_0^{s'} \ln(\bar{\gamma}(s''))\di s''}/\bar{t}(s')$. 
The constant $\kappa$ of a hyperbolic surface is recovered when $t_{R,n}$ and $t_{L,n}$ are constants.  
Changing $t_{R,n}$ and $t_{L,n}$ then tunes local curvatures. 
For instance, when $t_{R,n}=\frac{\hbar^2}{2Md^2}e^{-\Theta(n-n^*)/(2n)}$, $t_{L,n}=\frac{\hbar^2}{2Md^2}e^{\Theta(n-n^*)/(2n)}$, where $\Theta(x)$ is the Heaviside step function, the curvature vanishes everywhere except at a particular location, i.e., $\kappa\sim \delta(y-y^*)$, where $y^*=y_0+ n^*d$. 

Our scheme can be straightforwardly generalized to higher dimensions. 
For instance, 1D HN chains can be assembled to access higher dimensional curved spaces and inter-chain couplings may fundamentally change the curvatures (Supplementary Materials).

\begin{figure}[t] 
  \includegraphics [angle=0,width=.48\textwidth]
  {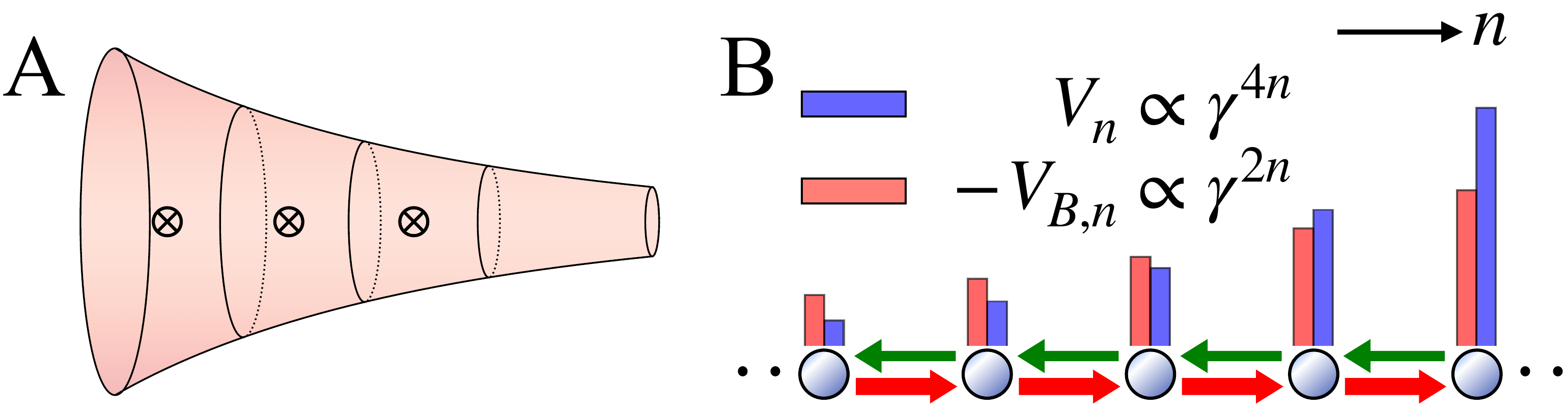}
  \caption
  {(A)A hyperbolic surface threaded by uniform magnetic fluxes.
   (B)An extra onsite energy in HN chain, $V_{B,n}$, encapsulates the magnetic field.
  }\label{fig:fig4}
\end{figure}

The duality we established has a wide range of profound applications. 
For instance, Fig.~\ref{fig:fig4}A shows a non-Hermitian realization of QHS in curved spaces. 
When a particle with a charge $-e$ is subjected to a uniform magnetic field, $y^2\nabla^2$ in Eq.(\ref{PH}) is replaced by $y^2[(\partial_x-i\frac{eB}{\hbar \kappa}\frac1{y})^2+\partial_y^2]$, where  we have chosen the gauge with the vector potential $\vec{A}=(-B/(\kappa y),0)$ \cite{Comtet1987} such that $k_x$ is still a good quantum number.
Wavefunctions of the lowest Landau level (LLL) are written as,
\begin{equation}
\psi_{\rm LLL}=(2k_x)^{\frac{eB}{\hbar\kappa}-\frac{1}{2}}\sqrt{\frac{\mathcal{N}\kappa}{\Gamma(2\frac{eB}{\hbar\kappa}-1)L}}e^{-k_x y+ik_xx}y^{\frac{eB}{\hbar\kappa}},\label{LLL}
\end{equation}
whose eigenenergies, $E_{LLL}=-\frac{\hbar^2\kappa}{8M}+\frac{\hbar eB}{2M}$, are independent of $k_x$, manifesting the degeneracy of the Landau levels.  
In the dual non-Hermitian systems, a finite magnetic field corresponds to an extra onsite potential in Eq.(\ref{LM}), $V_n\rightarrow V_n+V_{B,n}$, where  $V_{B,n}=-ba\gamma^{2n} \sqrt{t_Lt_R}$. 
The dimensionless $b$ characterizing the strength of $V_{B,n}$ relative to $V_n$ is written as
\begin{equation}
  b={eBd^2}/(\hbar \ln|\gamma|),
\end{equation}
where $y_0k_x=a(2\ln|\gamma|)^{-1}$ has been used. 
The magnetic flux density, $\rho_\phi=eB/(2\pi \hbar)=\frac{\ln|\gamma|}{2\pi d^2}b$, is thus determined by the ratio of $V_{B,n}$ to $V_n$. 

A complete description of QHS requires its gravitational responses in curved spaces. 
For instance, the particle density, $\rho$, depends on the local curvature~\cite{Wen1992}, 
\begin{equation}
\rho=\nu\rho_\phi-\kappa/(4\pi), \label{dencurve}
\end{equation} 
where $\nu$ is the filling factor. 
For integer QHS, $\nu=1$ and Eq.~(\ref{dencurve}) for a hyperbolic surface can be straightforwardly proved using Eq.(\ref{LLL}) (Supplementary Materials). 
The counterpart of Eq.(\ref{dencurve}) in the non-Hermitian lattice is
\begin{equation}
  |\gamma|^{2n}\int \di a\mathcal{N}_n(a,b)
  =b\ln(|\gamma|)-2\ln^2(|\gamma|), \label{dennon}
\end{equation} 
where $\mathcal{N}_n(a,b)=|\gamma|^{-2n}\psi_n^*\psi_n$ is the particle number at lattice site $n$ in the non-Hermitian system.
Mapping the magnetic flux, $eB/(2\pi\hbar)$, to the ratio between onsite potentials, $b\ln(|\gamma|)/(2\pi d^2)$, 
Eq.(\ref{dencurve}) and Eq.(\ref{dennon}) are equivalent.
The dependence of densities of QHS on curvatures is thus readily detectable using this non-Hermitian realization.
An alternative scheme is to implement a 2D non-Hermitian lattice model, which serves as a non-Hermitian generalization of the Harper-Hofstadter Hamiltonian~\cite{Hofstadter1976} (Supplementary Materials).

In parallel to accessing curved spaces using non-Hermitian systems, experimentalists could also use curved spaces to study non-Hermitian physics~\cite{Bekenstein2017,ZhouZW2018,Kollr2019}. 
In conventional understandings, non-Hermiticity arises when dissipations exist. 
While dissipations have been engineered in certain apparatuses to deliver desired non-Hermitian Hamiltonians~\cite{Syassen2008,LuoL2019,Yan2020}, in other platforms, such engineering might be more difficult and sometime experimentalists may have to use indirect means such as simulating non-Hermitian quantum walks~\cite{Regensburger2012,Weidemann2020,Xiao2020}. 
Our results show that many non-Hermitian Hamiltonians are readily accessible using existing curved spaces. 
For instance, hyperbolic surfaces that have been created in laboratories could be used to realize the HN model directly. 
In particular, in contrast to current schemes used in the study of non-Hermitian physics, this method does not require engineering losses or gains. 
It thus provides a conceptually different protocol to access non-Hermiticity without dissipations. 

The duality we have found provides new perspectives for both the studies of non-Hermitian physics and curved spaces. 
As the curvature depends on the non-Hermiticity even when the separation between any two points in the system is not physically distorted, our conventional understandings of distance may need to be reformed. 
We hope that our work will stimulate more interests in studying deep connections between non-Hermitian physics and curved spaces.

{\bf Acknowledgements:}
QZ acknowledges useful discussions with Mahdi Hosseini and Pramey Upadhyaya about realizations of chiral couplings. 
Q. Zhou and C. Lv are supported by the Air Force Office of Scientific Research under award number FA9550-20-1-0221, DOE DE-SC0019202, 
DOE QuantISED program of the theory consortium ``Intersections of QIS and Theoretical Particle Physics" at Fermilab,
W. M. Keck Foundation, 
and a seed grant from PQSEI. R. Zhang is supported by the National Key R$\&$D Program of China (Grant No. 2018YFA0307601), NSFC (Grant No.12174300, 11804268).
{\bf Author Contributions:} Q. Z. conceived the idea. C. Lv,  R. Z. and Q. Z. performed analytical and numerical calculations with inputs from Z. Z. on parts of finite magnetic fields.  Q. Z. wrote the manuscript with inputs from all authors.
{\bf Additional Information:}
The authors declare no competing financial interests. 
Supplementary Information is available for this paper. 
Correspondence and requests for materials should be addressed to Q.Z.


%
\onecolumngrid
\newpage
\vspace{0.4in}
\centerline{\bf\large Supplementary Materials for ``Curving the space by non-Hermiticity"}
\setcounter{equation}{0}
\setcounter{figure}{0}
\setcounter{table}{0}
\makeatletter
\renewcommand{\theequation}{S\arabic{equation}}
\renewcommand{\thefigure}{S\arabic{figure}}
\renewcommand{\thetable}{S\arabic{table}}
\vspace{0.2in}
\subsection{Ricci scalar curvature and mean curvature}

In the main text, we considered an intrinsic curved space where the quantum particle couples to the Ricci scalar curvature $R_{\rm Ric}$. 
Whereas the choice of the coupling constant is not unique~\cite{SCarroll2019}, we have adopted the one used in Ref.~\cite{SGutzwiller1983}.
In 2D, the Ricci scalar curvature relates to the Gaussian curvature $-\kappa$ through $-\kappa=R_{\rm Ric}/2$.
We therefore obtain a potential term proportional to the Gaussian curvature $-\kappa$.

In addition to intrinsic curved spaces, an alternative approach is to consider curved spaces embedded in a higher dimensional flat space.  
This requires applying physical constraints in the higher dimensional space such that particles can only move in a certain subspace that is curved.  
The physical constraints thus induce an extra potential that also depends on the mean curvature~\cite{SCosta1982}.
Considering the pseudosphere, which is an embedded hyperbolic surface in 3D as shown in Eq.~(9) of the main text, its first and second fundamental forms are written as 
\begin{equation}
   {\bf g}=\frac1{\kappa}\begin{pmatrix}\tanh^2(\eta)&0\\0&\sech^2(\eta)\end{pmatrix},\quad 
   {\bf h}=\frac1{\sqrt{\kappa}}\begin{pmatrix}-\sech(\eta)\tanh(\eta)&0\\0&\sech(\eta)\tanh(\eta)\end{pmatrix}, 
\end{equation}
respectively. The mean curvature and Gaussian curvature become 
\begin{equation}
   \begin{split}
      &K_{\rm mean}=\frac1{2g}(g_{11}h_{22}+g_{22}h_{11}-2g_{12}h_{12})=\frac1{4}\sqrt{\kappa}(-3 + \cosh(2 \eta))\csch(\eta),\\
      &K_{\rm Gaussian}=\frac1{g}\det({\bf h})=-\kappa,
   \end{split}
\end{equation}
respectively. 
The \Schrodinger equation that describes quantum particles confined to this hyperbolic surface through an external potential along the normal direction contains a surface potential~\cite{SCosta1982}
\begin{equation}
   V_S(\eta)=-\frac{\hbar^2}{2M}\left(K^2_{\rm mean}-K_{\rm Gaussian}\right)=
   -\frac{\hbar^2\kappa}{2M}\left(\frac1{16}(-3+\cosh(2 \eta))^2\csch^2(\eta)+1\right).
\end{equation} 
Transforming to the $s$-coordinate, it amounts to 
\begin{equation}
   V_S(s)=-\frac{\hbar^2\kappa}{8M}\frac{e^{4 \sqrt{\kappa} s}}{e^{2 \sqrt{\kappa} s}-1}.
\end{equation}
We have chosen the constant $r_0=y_0$ such that $\eta(s=0)=0$. $ V_S$ only contributes to an extra potential in the curved space once applying our duality.
Correspondingly, an on-site potential $V_{S,n}=V_S(nd)$ introduced in the non-Hermitian model allows us to simulate embedded curved spaces.

\subsection{The HN model with the next-nearest-neighbor hopping}

In the presence of next-nearest-neighbor hoppings, the non-Hermitian model is written as 
\begin{equation}
   -t_{R2}\psi_{n-2}-t_{L2}\psi_{n+2}-t_{R1}\psi_{n-1}-t_{L1}\psi_{n+1}=E\psi_n. \label{HNt2}
\end{equation}
A generic solution to this \Schrodinger equation is written as $\beta^n$,
with eigenenergy $E(\beta)=-(t_{R2} \beta^{-2}+t_{L2} \beta^{2}+t_{R1}\beta^{-1}+t_{L1}\beta)$. 
For the open boundar y condition, $\psi_{n<0}=\psi_{n>N-1}=0$, 
the \Schrodinger equation formally becomes different near the edges,
\begin{equation}
   \begin{split}
      -t_{L2}\psi_{2}-t_{L1}\psi_{1}=E\psi_0,&\quad -t_{L2}\psi_{3}-t_{R1}\psi_{0}-t_{L1}\psi_{2}=E\psi_1,\\
      -t_{R2}\psi_{N-3}-t_{R1}\psi_{N-2}=E\psi_{N-1},&\quad -t_{R2}\psi_{N-4}-t_{R1}\psi_{N-3}-t_{L1}\psi_{N-1}=E\psi_{N-2}.\label{HNt2Bc}
   \end{split}
\end{equation}
Alternatively, the boundary condition can be written as $\psi_{-1}=\psi_{-2}=\psi_{N}=\psi_{N+1}=0$.

\begin{figure}[b]
       \centering
       \includegraphics[width=.5\textwidth]{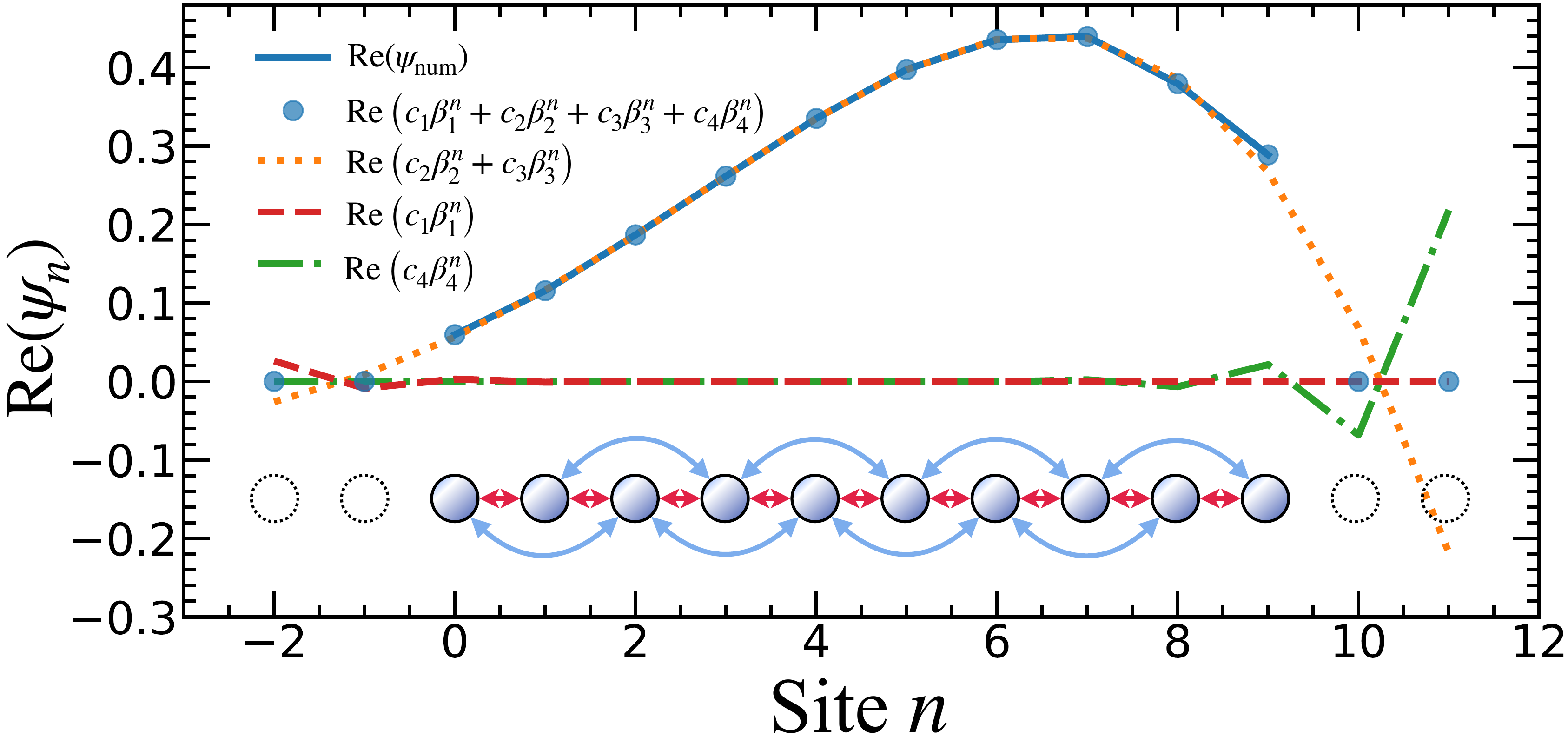}
       \caption{An eigenstate of the HN model with the next nearest neighbor tunnelings. We use $N=10$, $t_{L1}=0.5t_{R1}$, $t_{R2}=0.6t_{R1}$, and $t_{L2}=0.4t_{R1}$.
       The real parts of wavefunction are shown and  the imaginary parts vanish.
       The numerical solution is extrapolated to the lattice sites $-1$, $-2$, $10$ and $11$ by $\psi_{\rm num}=c_1\beta_1^n+c_2\beta_2^n+c_3\beta_3^n+c_4\beta_4^n$.
       Here, $c_2\beta_2^n+c_3\beta_3^n$ forms a standing wave, $c_1\beta_1^n$ and $c_4\beta_4^n$ are localized at the boundary to fulfill
       the boundary condition.}
       \label{fig:figs1} 
\end{figure}

For a given complex energy $E$, $t_{L2} \beta^{4}+t_{L1}\beta^3+E\beta^{2}+t_{R1}\beta+t_{R2}=0$ has four solutions, 
denoted by $|\beta_1|\leq|\beta_2|\leq|\beta_3|\leq|\beta_4|$.
A specific solution to boundary conditions Eq.~(\ref{HNt2Bc}) is written as 
\begin{equation}
   \psi_n=c_1\beta_1^n+c_2\beta_2^n+c_3\beta_3^n+c_4\beta_4^n,
\end{equation}
It is known that $|\beta_1|\leq|\beta_2|=|\beta_3|\leq|\beta_4|$ 
is required to satisfy the boundary condition when $N\to\infty$. This leads to the definition of the Generalized Brillouin Zone (GBZ)~\cite{SWang2018,SYokomizo2019,SHu2020}.  
As such, $\beta_2^n$ and $\beta_3^n$ are dominant in the bulk. An eigenstate is depicted in Fig.~\ref{fig:figs1}.

Using $\beta_2$ and $\beta_3$, we could formulate the effective theory in the bulk. 
We denote $\beta_2$ by $e^{(q+iK_0)d}$ and $\beta_3$ by $e^{(q+iK_1)d}$
where $q$, $K_0$ and $K_1$ are real. 
Using the same method discussed in the main text, 
we define $\sqrt{d}\psi(s_n)\equiv\psi_n$, $\psi_0(s)=\phi_0(s)e^{iK_0s}e^{qs}$ such that $\phi_0(s)$ is slowly varying with changing $s$, 
the \Schrodinger equation for $\psi_0$ in the continuum limit is written as 
\begin{equation}
   -\mathcal{B}(K_0)\left[\partial_{s}^{2}
   -2\mathcal{A}(K_0)\partial_{s}
   +\mathcal{C}(K_0)\right]\psi_0(s)=E\psi_0(s),\label{nnnc}
\end{equation}
where 
\begin{equation}
   \begin{split}
      \mathcal{B}(K_0)&=\sum_{n=1,2}\left(t_{Rn}\beta_2^{-n}+t_{Ln}\beta_2^{n}\right)\frac{n^2d^2}{2},\\
      \mathcal{A}(K_0)&=(iK_0+q)-\frac{\sum_{n=1,2}\left(-t_{Rn}\beta_2^{-n}+t_{Ln}\beta_2^{n}\right)n}
      {d\sum_{n=1,2}\left(t_{Rn}\beta_2^{-n}+t_{Ln}\beta_2^{n}\right)n^2},\\
      \mathcal{C}(K_0)&=(iK_0+q)^2-(iK_0+q)\frac{2\sum_{n=1,2}\left(-t_{Rn}\beta_2^{-n}+t_{Ln}\beta_2^{n}\right)n}
      {d\sum_{n=1,2}\left(t_{Rn}\beta_2^{-n}+t_{Ln}\beta_2^{n}\right)n^2}+\frac{2\sum_{n=1,2}\left(t_{Rn}\beta_2^{-n}+t_{Ln}\beta_2^{n}\right)}
      {d^2\sum_{n=1,2}\left(t_{Rn}\beta_2^{-n}+t_{Ln}\beta_2^{n}\right)n^2}.
   \end{split}
\end{equation}

Since $q$ has been determined by $K_0$ in GBZ, $\mathcal{A}(K_0)$, $\mathcal{B}(K_0)$, $\mathcal{C}(K_0)$ are known. 
A coordinate transformation $y/y_0=e^{2s \mathcal{A}_R}$ leads to 
\begin{equation}
   -\mathcal{B}(K_0)\left[4\mathcal{A}_R(K_0)^2 y^2\partial_{y}^{2}
   -4i\mathcal{A}_I(K_0)\mathcal{A}_R(K_0) y\partial_{y}
   +\mathcal{C}(K_0)\right]\psi_0(y)=E\psi_0(y),\label{nnnc2php}
\end{equation}
where $\mathcal{A}_{R}$ and $\mathcal{A}_{I}$ denotes the real and imaginary part of $\mathcal{A}$, respectively.
Similarly, using $\beta_3$, we can define $\psi_1(s)=\phi_1(s)e^{iK_1s}e^{qs}$ and obtain, 
\begin{equation}
   -\mathcal{B}(K_1)\left[4\mathcal{A}_R(K_1)^2 y^2\partial_{y}^{2}
   -4i\mathcal{A}_I(K_1)\mathcal{A}_R(K_1) y\partial_{y}
   +\mathcal{C}(K_1)\right]\psi_1(y)=E\psi_1(y).\label{nnnc2php2}
\end{equation}
Both equations describe a Poincar\'e half-plane, 
which is subject to a real constant vector potential, 
$\vec{A}_{j=0,1}\sim \mathcal{A}_I(K_j)$, if $\mathcal{A}_I(K_j)\neq 0$. $\mathcal{A}_R(K_j)$ determines the curvature,  $-\kappa_j=-4\mathcal{A}^2_R(K_j)$,
and the mass is given by $\frac{\hbar^2}{2M}=\mathcal{B}(K_j)$.

When $t_{L2}$ and $t_{R2}$ vanish, we obtain $qd=\ln(\sqrt{t_{R1}/t_{L1}})$, and Eq.(\ref{nnnc2php}) reduces to, 
\begin{equation}
   \begin{split}
      -\sqrt{t_{L1}t_{R1}}\cos(K_0d)&\bigg[\left(2qdy\left(\partial_{y}-\frac1{2y}\right)-iK_0d\right)^2
      +4qd\tan(K_0d)iy\left(\partial_{y}-\frac1{2y}\right)+2K_0d\tan(K_0d)+2\bigg]\psi_0(y)=E\psi_0(y).
   \end{split}
\end{equation}
For $K_0d=0,\pi$, we recover the non-relativistic theory in the main text. $\mathcal{A}^2_R(K_0)$ also reduces to $q^2$ such that $-\kappa=-4q^2$, as expected.

For $K_0d\neq 0,\pi$, the first term in the square brackets of the above equation corresponds to high order corrections to the energy and the effective theory becomes
\begin{equation}
   \begin{split}
      -\sqrt{t_{L1}t_{R1}}\cos(K_0d)&\bigg[4qd\tan(K_0d)iy\left(\partial_{y}-\frac1{2y}\right)+2K_0d\tan(K_0d)+2\bigg]\psi_0(y)=E\psi_0(y).\label{nnnc2phpr1}
        \end{split}
\end{equation}
Similarly, we obtain 
\begin{equation}
   \begin{split}
        -\sqrt{t_{L1}t_{R1}}\cos(K_1d)&\bigg[4qd\tan(K_1d)iy\left(\partial_{y}-\frac1{2y}\right)+2K_1d\tan(K_1d)+2\bigg]\psi_1(y)=E\psi_1(y).\label{nnnc2phpr2}
   \end{split}
\end{equation}
Both equations describe effective theories for particles with linear dispersions on a \Poincare half-plane, 
$\hat{H}=-v_f\sqrt{\kappa}\frac1{2}(y\hat{p}_y+\hat{p}_yy)=i\hbar v_f\sqrt{\kappa}y\left(\partial_{y}-\frac1{2y}\right)$, 
where $v_f$ is the Fermi velocity and $\hat{p}_y=-i\hbar\left(\partial_{y}-\frac1{y}\right)$, $\hat{p}_x=-i\hbar\partial_{x}$. 
Comparing this Hamiltonian with that in Eq.~(\ref{nnnc2phpr1},\ref{nnnc2phpr2}), we obtain 
\begin{equation}
   -\kappa=-4q^2,\quad 
   v_{f,j}=-2\sqrt{t_{L1}t_{R1}}d\sin(K_jd)/\hbar ,\quad j=0,1,
\end{equation}
which recover the results in the main text, as expected.

\subsection{Coupled HN chains}
\begin{figure}[b]
   \includegraphics[angle=0,width=.95\textwidth]
   {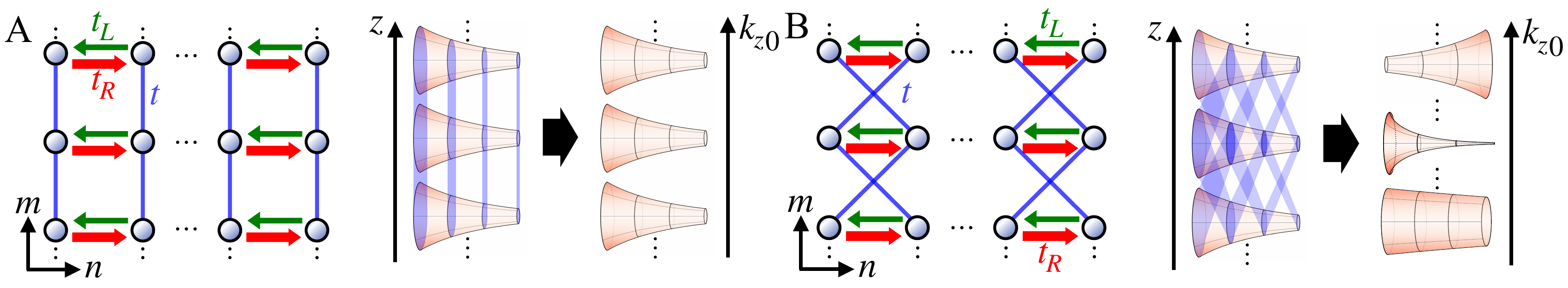}
   \caption
    {  A set of coupled HN chains is dual to a 3D curved space. (A) 
    The curvature of each decoupled surface remains unchanged by the vertical inter-chain couplings.
    (B) Inter-chain couplings fundamentally influence each curved surface, and the curvature becomes energy-dependent.
    }\label{fig:figs2}
\end{figure}
The lattice model 
with HN chains coupled vertically is illustrated in Fig.~\ref{fig:figs2}A, which reads 
\begin{equation}
  -t\Psi_{n,m+1}-t\Psi_{n,m-1}-t_R\Psi_{n-1,m}-t_L\Psi_{n+1,m}=E\Psi_{n,m}.\label{HNp1}
\end{equation} 
We define $\Psi(s,z)$ such that $\Psi(s_n,z_m)\equiv \Psi_{n,m}/d$, and  $\Psi(s,z)=\Phi(s,z)e^{qs}e^{iK_0s}e^{ik_{z0}z}$, where $\Phi$ is slowly varying. We have applied periodical boundary condition along the $z$-direction.
Near the band bottom $(K_0=0,k_{z0}=0)$, substituting $\Psi$ to Eq.~(\ref{HNp1}) and using the Taylor expansion of $\Phi(s,z)$,
we obtain
\begin{equation}
   \begin{split}
      -&\left[\left(2t+t_Re^{-qd}+t_Le^{qd}\right)+
      \left(-t_Re^{-qd}+t_Le^{qd}\right)d\partial_{s}+\frac{d^2}{2}\left(t_Re^{-qd}+t_Le^{qd}\right)\partial_{s}^{2}+
      td^2\partial_{z}\right]\Phi(s,z)=E\Phi(s,z)\label{HNp1c}
   \end{split}
\end{equation}
A generic solution is written as $\Phi=e^{ik_ss}e^{ik_zz}$. 
To ensure the open boundary condition in $s$-direction and obtain an effective theory near $(K_0=0,k_{z0}=0)$,
we require $E(k_s,0)=E(-k_s,0)$, which leads to $-t_Re^{-qd}+t_Le^{qd}=0$.
The solution gives $q=\ln(\sqrt{t_R/t_L})/d$. 
For a finite $k_z$, the solution to Eq.(\ref{HNp1c})
that satisfies the specified boundary conditions is written as $e^{ik_zz}\sin(k_ss)$. 
In other words, $q$ and the resultant curvature, is independent of $k_z$.  After a coordinate transformation $\frac{y}{y_0}=e^{2qs}$,
Eq.~(\ref{HNp1c}) is written as 
\begin{equation}
   \begin{split}
      -&\left[2\left(t+\sqrt{t_Lt_R}\right)
      +\sqrt{t_Lt_R} d^2 \kappa\left(y^2\partial_{y}^{2}+\frac{1}{4}\right)+
      td^2\partial_{z}^{2}\right]\Psi(y,z)=E\Psi(y,z),
   \end{split}
\end{equation}
where $\kappa=4\ln^2\left(\sqrt{\frac{t_R}{t_L}}\right)/d^2$.

In contrast, the model in Fig.~\ref{fig:figs2}B reads 
\begin{equation}
   -t\Psi_{n-1,m-1}-t\Psi_{n+1,m-1}-t\Psi_{n-1,m+1}-t\Psi_{n+1,m+1}-t_R\Psi_{n-1,m}-t_L\Psi_{n+1,m}=E\Psi_{n,m}.\label{HNp2}
\end{equation} 
The slowly varying $\Phi(s,z)$ satisfies
\begin{equation}
   \begin{split}
      -&\left[\left(2te^{-qd}+2te^{qd}+t_Re^{-qd}+t_Le^{qd}\right)+
      \left(-2te^{-qd}+2te^{qd}-t_Re^{-qd}+t_Le^{qd}\right)d\partial_{s}\right.\\
      &\left.+\left(2te^{-qd}+2te^{qd}+t_Re^{-qd}+t_Le^{qd}\right)\frac1{2}d^2\partial_{s}^{2}+
      2t\left(e^{qd}+e^{-qd}\right)\frac1{2}d^2\partial_{z}^{2}
      +t\left(e^{qd}-e^{-qd}\right)d^3\partial_{z}^{2}\partial_{s}\right]\Phi(s,z)=E\Phi(s,z).
   \end{split}\label{HNp2c}
\end{equation}
For the effective theory near $(K_0=0,k_{z0}=0)$, the open boundary condition in $s$-direction requires
$E(k_s,0)=E(-k_s,0)$ and 
$-2te^{-qd}+2te^{qd}-t_Re^{-qd}+t_Le^{qd}=0$.
This provides us with $q=\frac1{d}\ln(\sqrt{\frac{t_R+2t}{t_L+2t}})$, and
\begin{equation}
   \begin{split}
      -&\left[2\tilde{t}+
      \tilde{t}d^2\partial_{s}^{2}+
      \frac{t}{\tilde{t}}(4t+t_L+t_R)d^2\partial_{z}^{2}
      +\frac{t}{\tilde{t}}(t_R-t_L)d^3\partial_{z}^{2}\partial_{s}\right]\Phi(s,z)=E\Phi(s,z),
   \end{split}\label{HNp2c2}
\end{equation}
where  $\tilde{t}=\sqrt{(t_R+2t)(t_L+2t)}$.
When $k_z$ is finite, the solution to Eq.(\ref{HNp2c2}) reads 
$\Phi=e^{ik_z z}\sin(k_s s)e^{\left(\frac{t(t_R-t_L)}{2\tilde{t}^2}k_zd\right)k_zs}$. 
The extra exponential function gives rise to a $k_z$-dependent 
$q(k_z)=\frac1{d}\ln(\sqrt{\frac{t_R+2t}{t_L+2t}})+\frac{td(t_R-t_L)}{2\tilde{t}^2}k_z^2$. After a coordinate transformation $\frac{y}{y_0}=e^{2qs}$ and $\Phi(s,z)=\Psi(s,z)e^{-qs}$, Eq.~(\ref{HNp2c2}) is written as 
\begin{equation}
   -\tilde{t}d^2\left[\frac2{d^2}+
   \kappa_c \left(y^2\partial_{y}^{2}+\frac1{4}\right)
   +\frac{t(t_R-t_L)}{\tilde{t}^2}\partial_{z}^{2}
   \left(\frac{4t+t_R+t_L}{t_R-t_L}+d\sqrt{\kappa_c}\left(y\partial_{y}-\frac1{2}\right)\right)\right]
   \Psi(y,z)=E\Psi(y,z), \label{HNp3}
\end{equation}
where $\kappa_c=\ln^2\left(\frac{t_R+2t}{t_L+2t}\right)/d^2$. 
A finite $k_z$ modifies the curvature, $\kappa_c(k_z)=\frac{1}{d^{2}}\ln^2\left(\frac{t_R+2t}{t_L+2t}\right)
+\frac{2t(t_R-t_L)}{\tilde{t}^2}\log(\frac{t_R+2t}{t_L+2t})k_z^2+{\cal O}((k_zd)^4)$.  

Whereas the above discussions apply to the band bottom, the effective theory can be formulated at any energies. 
In the lattice models, we write $\Psi_{n,m}=\psi_ne^{ik_{z0}md}$, and obtain 
\begin{equation}
   -2t\cos(k_{z0}d)\psi_n-t_R\psi_{n-1}-t_L\psi_{n+1}=E\psi_n,
 \end{equation} 
 and 
\begin{equation}
   -(t_R+2t\cos(k_{z0}d))\psi_{n-1}-(t_L+2t\cos(k_{z0}d))\psi_{n+1}=E\psi_{n},
\end{equation} 
for Eq.~(\ref{HNp1}) and (\ref{HNp2}), respectively.
Therefore, for each $k_{z0}$, we have a dual model in the curved space.
Especially, for Eq.~(\ref{HNp2}), we obtain a $k_{z0}$-dependent HN model and
the corresponding curved space can be derived in the same manner as in the main text, 
where $\kappa=\ln^2\left(\Big|\frac{t_R+2t\cos(k_{z0}d)}{t_L+2t\cos(k_{z0}d)}\Big|\right)/d^2$,
which is $k_{z0}$ dependent. It reduces to $\kappa_c$ if we take the limit of $k_{z0}d \to 0$.
 
\subsection{Gravitational responses of quantum Hall states in hyperbolic spaces}
The normalized wavefunction at the lowest Landau level on a \Poincare half-plane is written as 
\begin{equation}
   \Psi_{\rm LLL}(x,y)=2^{\alpha-1/2} k_x^{\alpha-1/2}\sqrt{\frac{\kappa\mathcal{N}}{\Gamma(2\alpha-1)L}}e^{ik_xx}e^{-k_xy}y^{\alpha},
\end{equation}
where $\Gamma(x)$ is the Gamma function $\alpha=\frac{eB}{\hbar\kappa}>0$, and $k_x=2\pi n/L$. The particle density at  unit filling becomes 
\begin{equation}
   \rho(x,y)=\frac1{\mathcal{N}}\sum_{k_x}|\Psi_{\rm LLL}|^2=\sum_{n=1}^\infty
    \left(\frac{4\pi n}{L}\right)^{2\alpha-1}e^{-\frac{4\pi n}{L}y}\frac{\kappa y^{2\alpha}}{\Gamma(2\alpha-1)L}=\left(\frac{4\pi}{L}\right)^{2\alpha-1}\frac{\kappa y^{2\alpha}}{\Gamma(2\alpha-1)L}\Li_{1-2\alpha}(e^{-\frac{4\pi}{L}y}),
\end{equation}
where $\Li_k(z)=\sum_{n=1}^\infty z^n/n^k$ is the polylogarithm function of order $k$. The summation is only taken from $n=1$ to $\infty$ since the wavefunction is not normalizable and Landau levels do not exist for $n\leq 0$. 
In the thermodynamic limit, 
\begin{equation}
   \lim_{L\to\infty}\rho(x,y)=\left(4\pi\right)^{-1}\frac{\Gamma(2\alpha)\kappa}{\Gamma(2\alpha-1)}
   =\frac{eB}{2\pi\hbar}-\frac{\kappa}{4\pi}.\label{grhp}
\end{equation}
We have used $\Li_k(e^w)=\Gamma(1-k)(-w)^{k-1}+\sum_{j=0}^\infty\frac{\zeta(k-j)}{j!}w^j$, 
which is valid for $|w|<2\pi$ and $k\notin \mathbb{N}^+$. $\zeta(x)$ denotes the zeta function.

\subsection{A non-Hermitian generalization of the Harper-Hofstadter model}

\begin{figure}[t] 
   \includegraphics [angle=0,width=.6\textwidth]
   {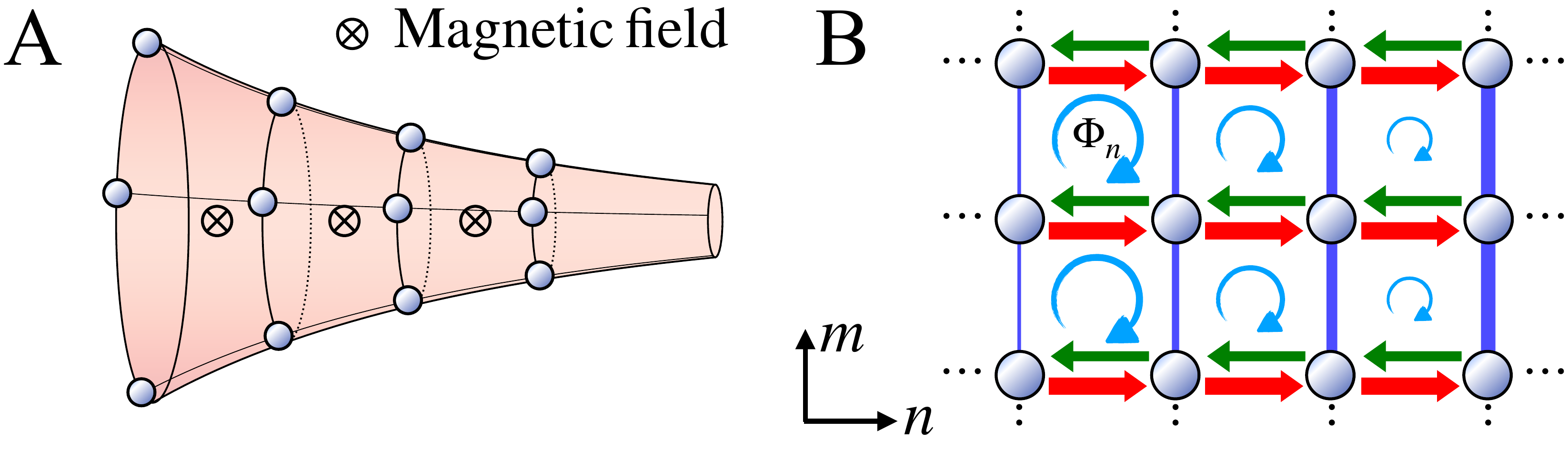}
   \caption
   {A realization of hyperbolic surface threaded by uniform magnetic fluxes using 2D non-Hermitian lattices. 
   Both vertical couplings and the magnetic flux per plaquette vary exponentially with changing $n$ since 
   the area of plaquettes changes exponentially on the hyperbolic surface. 
   }\label{fig:figs3}
 \end{figure}

 We consider the following 2D non-Hermitian lattice model,
\begin{equation}
\begin{split}
-t_R\Psi_{n-1,m}-t_L\Psi_{n+1,m}
-\tilde{a}^2\sqrt{t_Lt_R}\gamma^{4n}
\big( e^{i\theta_n}\Psi_{n,m-1}+e^{-i\theta_n}\Psi_{n,m+1}-2\Psi_{n,m}\big)=E\Psi_{n,m} \label{NHH}
\end{split}
\end{equation} 
where $\tilde{a}$ determines strengths of vertical tunnelings, $\theta_n=\tilde{b}\gamma^{-2n}/(2\tilde{a})$, and $\tilde{b}$ controls flux per plaquette.
As shown in Fig.~\ref{fig:figs3}B, Eq.(\ref{NHH}), a non-Hermitian generalization of the Harper-Hofstadter Hamiltonian~\cite{SHofstadter1976}, 
corresponds to discretizing a hyperbolic surface in both directions.
Eq.(\ref{NHH}) recovers the continuous model 
in the limit of small magnetic flux per plaquette, as we show below.
We write $\Psi_{n,m}=\psi_ne^{ik_{x}md}$, Eq.(\ref{NHH}) becomes 
\begin{equation}
     -t_R\psi_{n-1}-t_L\psi_{n+1}
     -2\tilde{a}^2\sqrt{t_Lt_R}\gamma^{4n}
     \bigg(-1+\cos(\frac{\tilde{b}}{2\tilde{a}}\gamma^{-2n}-k_xd)\bigg)\psi_{n}=E\psi_{n}.
\end{equation} 
The magnetic flux in each plaquette is proportional to $\frac{\tilde{b}}{\tilde{a}}$. 
When $k_xd\ll 1$ and $\frac{\tilde{b}}{\tilde{a}}\ll 1$, we obtain 
\begin{equation}
   -t_R\psi_{n-1}-t_L\psi_{n+1}
   +\tilde{a}^2\sqrt{t_Lt_R}
   \gamma^{4n}\bigg(\frac{\tilde{b}}{2\tilde{a}}\gamma^{-2n}-k_xd\bigg)^2\psi_{n}=E\psi_{n}.
\end{equation}
In the low-energy limit, the continuous model is written as 
\begin{equation}
   -\sqrt{t_Rt_L}d^2\left(\partial_{s}-\frac{\ln(\gamma)}{d}\right)^2\psi
   +\sqrt{t_Lt_R}d^2
   \tilde{a}^2 e^{4s\ln(\gamma)/d}\bigg(\frac{\tilde{b}}{2\tilde{a}d}e^{-2s\ln(\gamma)/d}-k_x\bigg)^2\psi=E\psi.
\end{equation}
Replacing
$k_x$ by $-i\partial_x$ and defining $\frac{y}{\tilde{a}d/(2\ln(\gamma))}=e^{2s\ln(\gamma)/d}$ yield the \Schrodinger equation of a charged particle 
on the \Poincare half-plane,
\begin{equation}
   -4\ln(\gamma)^2\sqrt{t_Lt_R}\left[y^2\left(\partial_{y}^{2}
   +\bigg(\partial_{x}-i\frac{\tilde{b}}{4\ln(\gamma) y}\bigg)^2\right)
   +\frac{1}{4}\right]\psi=E\psi.
\end{equation}
We can therefore identify the mass term 
$\frac{\hbar^2}{2M}=\sqrt{t_Lt_R}d^2$, 
curvature $-\kappa=-4\ln(\gamma)^2/d^2$, 
and the magnetic field $B=\frac{\hbar}{e}\frac{\tilde{b}\ln(\gamma)}{d^2}$. 
Adding interactions to this lattice model could potentially allow experimentalists to access fractional QHS with $\nu\neq 1$ or their counterparts of fractional Chern insulators in curved spaces.

\begin{thebibliography}{50}%
  \makeatletter
  \providecommand \@ifxundefined [1]{%
   \@ifx{#1\undefined}
  }%
  \providecommand \@ifnum [1]{%
   \ifnum #1\expandafter \@firstoftwo
   \else \expandafter \@secondoftwo
   \fi
  }%
  \providecommand \@ifx [1]{%
   \ifx #1\expandafter \@firstoftwo
   \else \expandafter \@secondoftwo
   \fi
  }%
  \providecommand \natexlab [1]{#1}%
  \providecommand \enquote  [1]{#1}%
  \providecommand \bibnamefont  [1]{#1}%
  \providecommand \bibfnamefont [1]{#1}%
  \providecommand \citenamefont [1]{#1}%
  \providecommand \href@noop [0]{\@secondoftwo}%
  \providecommand \href [0]{\begingroup \@sanitize@url \@href}%
  \providecommand \@href[1]{\@@startlink{#1}\@@href}%
  \providecommand \@@href[1]{\endgroup#1\@@endlink}%
  \providecommand \@sanitize@url [0]{\catcode `\\12\catcode `\$12\catcode
    `\&12\catcode `\#12\catcode `\^12\catcode `\_12\catcode `\%12\relax}%
  \providecommand \@@startlink[1]{}%
  \providecommand \@@endlink[0]{}%
  \providecommand \url  [0]{\begingroup\@sanitize@url \@url }%
  \providecommand \@url [1]{\endgroup\@href {#1}{\urlprefix }}%
  \providecommand \urlprefix  [0]{URL }%
  \providecommand \Eprint [0]{\href }%
  \providecommand \doibase [0]{https://dx.doi.org}%
  \providecommand \selectlanguage [0]{\@gobble}%
  \providecommand \bibinfo  [0]{\@secondoftwo}%
  \providecommand \bibfield  [0]{\@secondoftwo}%
  \providecommand \translation [1]{[#1]}%
  \providecommand \BibitemOpen [0]{}%
  \providecommand \bibitemStop [0]{}%
  \providecommand \bibitemNoStop [0]{.\EOS\space}%
  \providecommand \EOS [0]{\spacefactor3000\relax}%
  \providecommand \BibitemShut  [1]{\csname bibitem#1\endcsname}%
  \let\auto@bib@innerbib\@empty
  \bibitem [{\citenamefont {Syassen}\ \emph {et~al.}(2008)\citenamefont
    {Syassen}, \citenamefont {Bauer}, \citenamefont {Lettner}, \citenamefont
    {Volz}, \citenamefont {Dietze}, \citenamefont {Garc{\'\i}a-Ripoll},
    \citenamefont {Cirac}, \citenamefont {Rempe},\ and\ \citenamefont
    {D{\"u}rr}}]{Syassen2008}%
    \BibitemOpen
    \bibfield  {author} {\bibinfo {author} {\bibfnamefont {N.}~\bibnamefont
    {Syassen}}, \bibinfo {author} {\bibfnamefont {D.~M.}\ \bibnamefont {Bauer}},
    \bibinfo {author} {\bibfnamefont {M.}~\bibnamefont {Lettner}}, \bibinfo
    {author} {\bibfnamefont {T.}~\bibnamefont {Volz}}, \bibinfo {author}
    {\bibfnamefont {D.}~\bibnamefont {Dietze}}, \bibinfo {author} {\bibfnamefont
    {J.~J.}\ \bibnamefont {Garc{\'\i}a-Ripoll}}, \bibinfo {author} {\bibfnamefont
    {J.~I.}\ \bibnamefont {Cirac}}, \bibinfo {author} {\bibfnamefont
    {G.}~\bibnamefont {Rempe}}, \ and\ \bibinfo {author} {\bibfnamefont
    {S.}~\bibnamefont {D{\"u}rr}},\ }\bibfield  {title} {\bibinfo {title} {Strong
    dissipation inhibits losses and induces correlations in cold molecular
    gases},\ }\href {\doibase/10.1126/science.1155309} {\bibfield  {journal}
    {\bibinfo  {journal} {Science}\ }\textbf {\bibinfo {volume} {320}},\ \bibinfo
    {pages} {1329} (\bibinfo {year} {2008})}\BibitemShut {NoStop}%
  \bibitem [{\citenamefont {Regensburger}\ \emph {et~al.}(2012)\citenamefont
    {Regensburger}, \citenamefont {Bersch}, \citenamefont {Miri}, \citenamefont
    {Onishchukov}, \citenamefont {Christodoulides},\ and\ \citenamefont
    {Peschel}}]{Regensburger2012}%
    \BibitemOpen
    \bibfield  {author} {\bibinfo {author} {\bibfnamefont {A.}~\bibnamefont
    {Regensburger}}, \bibinfo {author} {\bibfnamefont {C.}~\bibnamefont
    {Bersch}}, \bibinfo {author} {\bibfnamefont {M.-A.}\ \bibnamefont {Miri}},
    \bibinfo {author} {\bibfnamefont {G.}~\bibnamefont {Onishchukov}}, \bibinfo
    {author} {\bibfnamefont {D.~N.}\ \bibnamefont {Christodoulides}}, \ and\
    \bibinfo {author} {\bibfnamefont {U.}~\bibnamefont {Peschel}},\ }\bibfield
    {title} {\bibinfo {title} {Parity{\textendash}time synthetic photonic
    lattices},\ }\href {\doibase/10.1038/nature11298} {\bibfield  {journal}
    {\bibinfo  {journal} {Nature}\ }\textbf {\bibinfo {volume} {488}},\ \bibinfo
    {pages} {167} (\bibinfo {year} {2012})}\BibitemShut {NoStop}%
  \bibitem [{\citenamefont {Feng}\ \emph {et~al.}(2014)\citenamefont {Feng},
    \citenamefont {Wong}, \citenamefont {Ma}, \citenamefont {Wang},\ and\
    \citenamefont {Zhang}}]{Feng2014}%
    \BibitemOpen
    \bibfield  {author} {\bibinfo {author} {\bibfnamefont {L.}~\bibnamefont
    {Feng}}, \bibinfo {author} {\bibfnamefont {Z.~J.}\ \bibnamefont {Wong}},
    \bibinfo {author} {\bibfnamefont {R.-M.}\ \bibnamefont {Ma}}, \bibinfo
    {author} {\bibfnamefont {Y.}~\bibnamefont {Wang}}, \ and\ \bibinfo {author}
    {\bibfnamefont {X.}~\bibnamefont {Zhang}},\ }\bibfield  {title} {\bibinfo
    {title} {Single-mode laser by parity-time symmetry breaking},\ }\href
    {\doibase/10.1126/science.1258479} {\bibfield  {journal} {\bibinfo  {journal}
    {Science}\ }\textbf {\bibinfo {volume} {346}},\ \bibinfo {pages} {972}
    (\bibinfo {year} {2014})}\BibitemShut {NoStop}%
  \bibitem [{\citenamefont {Longhi}\ \emph {et~al.}(2015)\citenamefont {Longhi},
    \citenamefont {Gatti},\ and\ \citenamefont {Valle}}]{Longhi2015}%
    \BibitemOpen
    \bibfield  {author} {\bibinfo {author} {\bibfnamefont {S.}~\bibnamefont
    {Longhi}}, \bibinfo {author} {\bibfnamefont {D.}~\bibnamefont {Gatti}}, \
    and\ \bibinfo {author} {\bibfnamefont {G.~D.}\ \bibnamefont {Valle}},\
    }\bibfield  {title} {\bibinfo {title} {Robust light transport in
    non-Hermitian photonic lattices},\ }\href {\doibase/10.1038/srep13376}
    {\bibfield  {journal} {\bibinfo  {journal} {Scientific Reports}\ }\textbf
    {\bibinfo {volume} {5}} (\bibinfo {year} {2015}),\
    10.1038/srep13376}\BibitemShut {NoStop}%
  \bibitem [{\citenamefont {Chen}\ \emph {et~al.}(2017)\citenamefont {Chen},
    \citenamefont {\"{O}zdemir}, \citenamefont {Zhao}, \citenamefont {Wiersig},\
    and\ \citenamefont {Yang}}]{Chen2017}%
    \BibitemOpen
    \bibfield  {author} {\bibinfo {author} {\bibfnamefont {W.}~\bibnamefont
    {Chen}}, \bibinfo {author} {\bibfnamefont {{\c{S}}.~K.}\ \bibnamefont
    {\"{O}zdemir}}, \bibinfo {author} {\bibfnamefont {G.}~\bibnamefont {Zhao}},
    \bibinfo {author} {\bibfnamefont {J.}~\bibnamefont {Wiersig}}, \ and\
    \bibinfo {author} {\bibfnamefont {L.}~\bibnamefont {Yang}},\ }\bibfield
    {title} {\bibinfo {title} {Exceptional points enhance sensing in an optical
    microcavity},\ }\href {\doibase/10.1038/nature23281} {\bibfield  {journal}
    {\bibinfo  {journal} {Nature}\ }\textbf {\bibinfo {volume} {548}},\ \bibinfo
    {pages} {192} (\bibinfo {year} {2017})}\BibitemShut {NoStop}%
  \bibitem [{\citenamefont {Li}\ \emph {et~al.}(2019)\citenamefont {Li},
    \citenamefont {Harter}, \citenamefont {Liu}, \citenamefont {de~Melo},
    \citenamefont {Joglekar},\ and\ \citenamefont {Luo}}]{LuoL2019}%
    \BibitemOpen
    \bibfield  {author} {\bibinfo {author} {\bibfnamefont {J.}~\bibnamefont
    {Li}}, \bibinfo {author} {\bibfnamefont {A.~K.}\ \bibnamefont {Harter}},
    \bibinfo {author} {\bibfnamefont {J.}~\bibnamefont {Liu}}, \bibinfo {author}
    {\bibfnamefont {L.}~\bibnamefont {de~Melo}}, \bibinfo {author} {\bibfnamefont
    {Y.~N.}\ \bibnamefont {Joglekar}}, \ and\ \bibinfo {author} {\bibfnamefont
    {L.}~\bibnamefont {Luo}},\ }\bibfield  {title} {\bibinfo {title} {Observation
    of parity-time symmetry breaking transitions in a dissipative Floquet system
    of ultracold atoms},\ }\href {\doibase/10.1038/s41467-019-08596-1} {\bibfield
     {journal} {\bibinfo  {journal} {Nature Communications}\ }\textbf {\bibinfo
    {volume} {10}} (\bibinfo {year} {2019}),\
    10.1038/s41467-019-08596-1}\BibitemShut {NoStop}%
  \bibitem [{\citenamefont {Gou}\ \emph {et~al.}(2020)\citenamefont {Gou},
    \citenamefont {Chen}, \citenamefont {Xie}, \citenamefont {Xiao},
    \citenamefont {Deng}, \citenamefont {Gadway}, \citenamefont {Yi},\ and\
    \citenamefont {Yan}}]{Yan2020}%
    \BibitemOpen
    \bibfield  {author} {\bibinfo {author} {\bibfnamefont {W.}~\bibnamefont
    {Gou}}, \bibinfo {author} {\bibfnamefont {T.}~\bibnamefont {Chen}}, \bibinfo
    {author} {\bibfnamefont {D.}~\bibnamefont {Xie}}, \bibinfo {author}
    {\bibfnamefont {T.}~\bibnamefont {Xiao}}, \bibinfo {author} {\bibfnamefont
    {T.-S.}\ \bibnamefont {Deng}}, \bibinfo {author} {\bibfnamefont
    {B.}~\bibnamefont {Gadway}}, \bibinfo {author} {\bibfnamefont
    {W.}~\bibnamefont {Yi}}, \ and\ \bibinfo {author} {\bibfnamefont
    {B.}~\bibnamefont {Yan}},\ }\bibfield  {title} {\bibinfo {title} {Tunable
    nonreciprocal quantum transport through a dissipative Aharonov-Bohm ring in
    ultracold atoms},\ }\href {\doibase/10.1103/PhysRevLett.124.070402}
    {\bibfield  {journal} {\bibinfo  {journal} {Phys. Rev. Lett.}\ }\textbf
    {\bibinfo {volume} {124}},\ \bibinfo {pages} {070402} (\bibinfo {year}
    {2020})}\BibitemShut {NoStop}%
  \bibitem [{\citenamefont {Weidemann}\ \emph {et~al.}(2020)\citenamefont
    {Weidemann}, \citenamefont {Kremer}, \citenamefont {Helbig}, \citenamefont
    {Hofmann}, \citenamefont {Stegmaier}, \citenamefont {Greiter}, \citenamefont
    {Thomale},\ and\ \citenamefont {Szameit}}]{Weidemann2020}%
    \BibitemOpen
    \bibfield  {author} {\bibinfo {author} {\bibfnamefont {S.}~\bibnamefont
    {Weidemann}}, \bibinfo {author} {\bibfnamefont {M.}~\bibnamefont {Kremer}},
    \bibinfo {author} {\bibfnamefont {T.}~\bibnamefont {Helbig}}, \bibinfo
    {author} {\bibfnamefont {T.}~\bibnamefont {Hofmann}}, \bibinfo {author}
    {\bibfnamefont {A.}~\bibnamefont {Stegmaier}}, \bibinfo {author}
    {\bibfnamefont {M.}~\bibnamefont {Greiter}}, \bibinfo {author} {\bibfnamefont
    {R.}~\bibnamefont {Thomale}}, \ and\ \bibinfo {author} {\bibfnamefont
    {A.}~\bibnamefont {Szameit}},\ }\bibfield  {title} {\bibinfo {title}
    {Topological funneling of light},\ }\href {\doibase/10.1126/science.aaz8727}
    {\bibfield  {journal} {\bibinfo  {journal} {Science}\ }\textbf {\bibinfo
    {volume} {368}},\ \bibinfo {pages} {311} (\bibinfo {year}
    {2020})}\BibitemShut {NoStop}%
  \bibitem [{\citenamefont {Helbig}\ \emph {et~al.}(2020)\citenamefont {Helbig},
    \citenamefont {Hofmann}, \citenamefont {Imhof}, \citenamefont {Abdelghany},
    \citenamefont {Kiessling}, \citenamefont {Molenkamp}, \citenamefont {Lee},
    \citenamefont {Szameit}, \citenamefont {Greiter},\ and\ \citenamefont
    {Thomale}}]{Helbig2020}%
    \BibitemOpen
    \bibfield  {author} {\bibinfo {author} {\bibfnamefont {T.}~\bibnamefont
    {Helbig}}, \bibinfo {author} {\bibfnamefont {T.}~\bibnamefont {Hofmann}},
    \bibinfo {author} {\bibfnamefont {S.}~\bibnamefont {Imhof}}, \bibinfo
    {author} {\bibfnamefont {M.}~\bibnamefont {Abdelghany}}, \bibinfo {author}
    {\bibfnamefont {T.}~\bibnamefont {Kiessling}}, \bibinfo {author}
    {\bibfnamefont {L.~W.}\ \bibnamefont {Molenkamp}}, \bibinfo {author}
    {\bibfnamefont {C.~H.}\ \bibnamefont {Lee}}, \bibinfo {author} {\bibfnamefont
    {A.}~\bibnamefont {Szameit}}, \bibinfo {author} {\bibfnamefont
    {M.}~\bibnamefont {Greiter}}, \ and\ \bibinfo {author} {\bibfnamefont
    {R.}~\bibnamefont {Thomale}},\ }\bibfield  {title} {\bibinfo {title}
    {Generalized bulk{\textendash}boundary correspondence in non-Hermitian
    topolectrical circuits},\ }\href {\doibase/10.1038/s41567-020-0922-9}
    {\bibfield  {journal} {\bibinfo  {journal} {Nature Physics}\ }\textbf
    {\bibinfo {volume} {16}},\ \bibinfo {pages} {747} (\bibinfo {year}
    {2020})}\BibitemShut {NoStop}%
  \bibitem [{\citenamefont {Yao}\ and\ \citenamefont {Wang}(2018)}]{Wang2018}%
    \BibitemOpen
    \bibfield  {author} {\bibinfo {author} {\bibfnamefont {S.}~\bibnamefont
    {Yao}}\ and\ \bibinfo {author} {\bibfnamefont {Z.}~\bibnamefont {Wang}},\
    }\bibfield  {title} {\bibinfo {title} {Edge states and topological invariants
    of non-Hermitian systems},\ }\href {\doibase/10.1103/PhysRevLett.121.086803}
    {\bibfield  {journal} {\bibinfo  {journal} {Phys. Rev. Lett.}\ }\textbf
    {\bibinfo {volume} {121}},\ \bibinfo {pages} {086803} (\bibinfo {year}
    {2018})}\BibitemShut {NoStop}%
  \bibitem [{\citenamefont {Kunst}\ \emph {et~al.}(2018)\citenamefont {Kunst},
    \citenamefont {Edvardsson}, \citenamefont {Budich},\ and\ \citenamefont
    {Bergholtz}}]{Kunst2018}%
    \BibitemOpen
    \bibfield  {author} {\bibinfo {author} {\bibfnamefont {F.~K.}\ \bibnamefont
    {Kunst}}, \bibinfo {author} {\bibfnamefont {E.}~\bibnamefont {Edvardsson}},
    \bibinfo {author} {\bibfnamefont {J.~C.}\ \bibnamefont {Budich}}, \ and\
    \bibinfo {author} {\bibfnamefont {E.~J.}\ \bibnamefont {Bergholtz}},\
    }\bibfield  {title} {\bibinfo {title} {Biorthogonal bulk-boundary
    correspondence in non-Hermitian systems},\ }\href
    {\doibase/10.1103/PhysRevLett.121.026808} {\bibfield  {journal} {\bibinfo
    {journal} {Phys. Rev. Lett.}\ }\textbf {\bibinfo {volume} {121}},\ \bibinfo
    {pages} {026808} (\bibinfo {year} {2018})}\BibitemShut {NoStop}%
  \bibitem [{\citenamefont {Martinez~Alvarez}\ \emph {et~al.}(2018)\citenamefont
    {Martinez~Alvarez}, \citenamefont {Barrios~Vargas},\ and\ \citenamefont
    {Foa~Torres}}]{Torres2018}%
    \BibitemOpen
    \bibfield  {author} {\bibinfo {author} {\bibfnamefont {V.~M.}\ \bibnamefont
    {Martinez~Alvarez}}, \bibinfo {author} {\bibfnamefont {J.~E.}\ \bibnamefont
    {Barrios~Vargas}}, \ and\ \bibinfo {author} {\bibfnamefont {L.~E.~F.}\
    \bibnamefont {Foa~Torres}},\ }\bibfield  {title} {\bibinfo {title}
    {Non-Hermitian robust edge states in one dimension: Anomalous localization
    and eigenspace condensation at exceptional points},\ }\href
    {\doibase/10.1103/PhysRevB.97.121401} {\bibfield  {journal} {\bibinfo
    {journal} {Phys. Rev. B}\ }\textbf {\bibinfo {volume} {97}},\ \bibinfo
    {pages} {121401} (\bibinfo {year} {2018})}\BibitemShut {NoStop}%
  \bibitem [{\citenamefont {Lee}\ and\ \citenamefont
    {Thomale}(2019)}]{Thomale2019}%
    \BibitemOpen
    \bibfield  {author} {\bibinfo {author} {\bibfnamefont {C.~H.}\ \bibnamefont
    {Lee}}\ and\ \bibinfo {author} {\bibfnamefont {R.}~\bibnamefont {Thomale}},\
    }\bibfield  {title} {\bibinfo {title} {Anatomy of skin modes and topology in
    non-Hermitian systems},\ }\href {\doibase/10.1103/PhysRevB.99.201103}
    {\bibfield  {journal} {\bibinfo  {journal} {Phys. Rev. B}\ }\textbf {\bibinfo
    {volume} {99}},\ \bibinfo {pages} {201103} (\bibinfo {year}
    {2019})}\BibitemShut {NoStop}%
  \bibitem [{\citenamefont {Borgnia}\ \emph {et~al.}(2020)\citenamefont
    {Borgnia}, \citenamefont {Kruchkov},\ and\ \citenamefont
    {Slager}}]{Slager2020}%
    \BibitemOpen
    \bibfield  {author} {\bibinfo {author} {\bibfnamefont {D.~S.}\ \bibnamefont
    {Borgnia}}, \bibinfo {author} {\bibfnamefont {A.~J.}\ \bibnamefont
    {Kruchkov}}, \ and\ \bibinfo {author} {\bibfnamefont {R.-J.}\ \bibnamefont
    {Slager}},\ }\bibfield  {title} {\bibinfo {title} {Non-Hermitian boundary
    modes and topology},\ }\href {\doibase/10.1103/PhysRevLett.124.056802}
    {\bibfield  {journal} {\bibinfo  {journal} {Phys. Rev. Lett.}\ }\textbf
    {\bibinfo {volume} {124}},\ \bibinfo {pages} {056802} (\bibinfo {year}
    {2020})}\BibitemShut {NoStop}%
  \bibitem [{\citenamefont {Zhang}\ \emph {et~al.}(2020)\citenamefont {Zhang},
    \citenamefont {Yang},\ and\ \citenamefont {Fang}}]{Fang2020}%
    \BibitemOpen
    \bibfield  {author} {\bibinfo {author} {\bibfnamefont {K.}~\bibnamefont
    {Zhang}}, \bibinfo {author} {\bibfnamefont {Z.}~\bibnamefont {Yang}}, \ and\
    \bibinfo {author} {\bibfnamefont {C.}~\bibnamefont {Fang}},\ }\bibfield
    {title} {\bibinfo {title} {Correspondence between winding numbers and skin
    modes in non-Hermitian systems},\ }\href
    {\doibase/10.1103/PhysRevLett.125.126402} {\bibfield  {journal} {\bibinfo
    {journal} {Phys. Rev. Lett.}\ }\textbf {\bibinfo {volume} {125}},\ \bibinfo
    {pages} {126402} (\bibinfo {year} {2020})}\BibitemShut {NoStop}%
  \bibitem [{\citenamefont {Okuma}\ \emph {et~al.}(2020)\citenamefont {Okuma},
    \citenamefont {Kawabata}, \citenamefont {Shiozaki},\ and\ \citenamefont
    {Sato}}]{Sato2020}%
    \BibitemOpen
    \bibfield  {author} {\bibinfo {author} {\bibfnamefont {N.}~\bibnamefont
    {Okuma}}, \bibinfo {author} {\bibfnamefont {K.}~\bibnamefont {Kawabata}},
    \bibinfo {author} {\bibfnamefont {K.}~\bibnamefont {Shiozaki}}, \ and\
    \bibinfo {author} {\bibfnamefont {M.}~\bibnamefont {Sato}},\ }\bibfield
    {title} {\bibinfo {title} {Topological origin of non-Hermitian skin
    effects},\ }\href {\doibase/10.1103/PhysRevLett.124.086801} {\bibfield
    {journal} {\bibinfo  {journal} {Phys. Rev. Lett.}\ }\textbf {\bibinfo
    {volume} {124}},\ \bibinfo {pages} {086801} (\bibinfo {year}
    {2020})}\BibitemShut {NoStop}%
  \bibitem [{\citenamefont {Bender}\ and\ \citenamefont
    {Boettcher}(1998)}]{Bender1998}%
    \BibitemOpen
    \bibfield  {author} {\bibinfo {author} {\bibfnamefont {C.~M.}\ \bibnamefont
    {Bender}}\ and\ \bibinfo {author} {\bibfnamefont {S.}~\bibnamefont
    {Boettcher}},\ }\bibfield  {title} {\bibinfo {title} {Real spectra in
    non-Hermitian Hamiltonians having $\textsc{P}\textsc{T}$ symmetry},\ }\href
    {\doibase/10.1103/PhysRevLett.80.5243} {\bibfield  {journal} {\bibinfo
    {journal} {Phys. Rev. Lett.}\ }\textbf {\bibinfo {volume} {80}},\ \bibinfo
    {pages} {5243} (\bibinfo {year} {1998})}\BibitemShut {NoStop}%
  \bibitem [{\citenamefont {Mostafazadeh}(2002)}]{Mostafazadeh2002}%
    \BibitemOpen
    \bibfield  {author} {\bibinfo {author} {\bibfnamefont {A.}~\bibnamefont
    {Mostafazadeh}},\ }\bibfield  {title} {\bibinfo {title} {Pseudo-hermiticity
    versus {PT} symmetry: The necessary condition for the reality of the spectrum
    of a non-Hermitian Hamiltonian},\ }\href {\doibase/10.1063/1.1418246}
    {\bibfield  {journal} {\bibinfo  {journal} {Journal of Mathematical Physics}\
    }\textbf {\bibinfo {volume} {43}},\ \bibinfo {pages} {205} (\bibinfo {year}
    {2002})}\BibitemShut {NoStop}%
  \bibitem [{\citenamefont {Ashida}\ \emph {et~al.}(2020)\citenamefont {Ashida},
    \citenamefont {Gong},\ and\ \citenamefont {Ueda}}]{Ashida2020}%
    \BibitemOpen
    \bibfield  {author} {\bibinfo {author} {\bibfnamefont {Y.}~\bibnamefont
    {Ashida}}, \bibinfo {author} {\bibfnamefont {Z.}~\bibnamefont {Gong}}, \ and\
    \bibinfo {author} {\bibfnamefont {M.}~\bibnamefont {Ueda}},\ }\bibfield
    {title} {\bibinfo {title} {Non-Hermitian physics},\ }\href
    {\doibase/10.1080/00018732.2021.1876991} {\bibfield  {journal} {\bibinfo
    {journal} {Advances in Physics}\ }\textbf {\bibinfo {volume} {69}},\ \bibinfo
    {pages} {249} (\bibinfo {year} {2020})}\BibitemShut {NoStop}%
  \bibitem [{\citenamefont {Bergholtz}\ \emph {et~al.}(2021)\citenamefont
    {Bergholtz}, \citenamefont {Budich},\ and\ \citenamefont
    {Kunst}}]{Bergholtz2021}%
    \BibitemOpen
    \bibfield  {author} {\bibinfo {author} {\bibfnamefont {E.~J.}\ \bibnamefont
    {Bergholtz}}, \bibinfo {author} {\bibfnamefont {J.~C.}\ \bibnamefont
    {Budich}}, \ and\ \bibinfo {author} {\bibfnamefont {F.~K.}\ \bibnamefont
    {Kunst}},\ }\bibfield  {title} {\bibinfo {title} {Exceptional topology of
    non-Hermitian systems},\ }\href {\doibase/10.1103/RevModPhys.93.015005}
    {\bibfield  {journal} {\bibinfo  {journal} {Rev. Mod. Phys.}\ }\textbf
    {\bibinfo {volume} {93}},\ \bibinfo {pages} {015005} (\bibinfo {year}
    {2021})}\BibitemShut {NoStop}%
  \bibitem [{\citenamefont {Xiong}(2018)}]{Xiong2018}%
    \BibitemOpen
    \bibfield  {author} {\bibinfo {author} {\bibfnamefont {Y.}~\bibnamefont
    {Xiong}},\ }\bibfield  {title} {\bibinfo {title} {Why does bulk boundary
    correspondence fail in some non-Hermitian topological models},\ }\href
    {\doibase/10.1088/2399-6528/aab64a} {\bibfield  {journal} {\bibinfo
    {journal} {Journal of Physics Communications}\ }\textbf {\bibinfo {volume}
    {2}},\ \bibinfo {pages} {035043} (\bibinfo {year} {2018})}\BibitemShut
    {NoStop}%
  \bibitem [{\citenamefont {Okuma}\ and\ \citenamefont {Sato}(2019)}]{Sato2019}%
    \BibitemOpen
    \bibfield  {author} {\bibinfo {author} {\bibfnamefont {N.}~\bibnamefont
    {Okuma}}\ and\ \bibinfo {author} {\bibfnamefont {M.}~\bibnamefont {Sato}},\
    }\bibfield  {title} {\bibinfo {title} {Topological phase transition driven by
    infinitesimal instability: Majorana fermions in non-Hermitian spintronics},\
    }\href {\doibase/10.1103/PhysRevLett.123.097701} {\bibfield  {journal}
    {\bibinfo  {journal} {Phys. Rev. Lett.}\ }\textbf {\bibinfo {volume} {123}},\
    \bibinfo {pages} {097701} (\bibinfo {year} {2019})}\BibitemShut {NoStop}%
  \bibitem [{\citenamefont {Wiersig}(2014)}]{Wiersig2014}%
    \BibitemOpen
    \bibfield  {author} {\bibinfo {author} {\bibfnamefont {J.}~\bibnamefont
    {Wiersig}},\ }\bibfield  {title} {\bibinfo {title} {Enhancing the sensitivity
    of frequency and energy splitting detection by using exceptional points:
    Application to microcavity sensors for single-particle detection},\ }\href
    {\doibase/10.1103/PhysRevLett.112.203901} {\bibfield  {journal} {\bibinfo
    {journal} {Phys. Rev. Lett.}\ }\textbf {\bibinfo {volume} {112}},\ \bibinfo
    {pages} {203901} (\bibinfo {year} {2014})}\BibitemShut {NoStop}%
  \bibitem [{\citenamefont {Kawabata}\ \emph {et~al.}(2019)\citenamefont
    {Kawabata}, \citenamefont {Shiozaki}, \citenamefont {Ueda},\ and\
    \citenamefont {Sato}}]{Kawabata2019}%
    \BibitemOpen
    \bibfield  {author} {\bibinfo {author} {\bibfnamefont {K.}~\bibnamefont
    {Kawabata}}, \bibinfo {author} {\bibfnamefont {K.}~\bibnamefont {Shiozaki}},
    \bibinfo {author} {\bibfnamefont {M.}~\bibnamefont {Ueda}}, \ and\ \bibinfo
    {author} {\bibfnamefont {M.}~\bibnamefont {Sato}},\ }\bibfield  {title}
    {\bibinfo {title} {Symmetry and topology in non-Hermitian physics},\ }\href
    {\doibase/10.1103/PhysRevX.9.041015} {\bibfield  {journal} {\bibinfo
    {journal} {Phys. Rev. X}\ }\textbf {\bibinfo {volume} {9}},\ \bibinfo {pages}
    {041015} (\bibinfo {year} {2019})}\BibitemShut {NoStop}%
  \bibitem [{\citenamefont {Lee}\ \emph {et~al.}(2019)\citenamefont {Lee},
    \citenamefont {Ahn}, \citenamefont {Zhou},\ and\ \citenamefont
    {Vishwanath}}]{LeeJY2019}%
    \BibitemOpen
    \bibfield  {author} {\bibinfo {author} {\bibfnamefont {J.~Y.}\ \bibnamefont
    {Lee}}, \bibinfo {author} {\bibfnamefont {J.}~\bibnamefont {Ahn}}, \bibinfo
    {author} {\bibfnamefont {H.}~\bibnamefont {Zhou}}, \ and\ \bibinfo {author}
    {\bibfnamefont {A.}~\bibnamefont {Vishwanath}},\ }\bibfield  {title}
    {\bibinfo {title} {Topological correspondence between Hermitian and
    non-Hermitian systems: Anomalous dynamics},\ }\href
    {\doibase/10.1103/PhysRevLett.123.206404} {\bibfield  {journal} {\bibinfo
    {journal} {Phys. Rev. Lett.}\ }\textbf {\bibinfo {volume} {123}},\ \bibinfo
    {pages} {206404} (\bibinfo {year} {2019})}\BibitemShut {NoStop}%
  \bibitem [{\citenamefont {Zhou}\ and\ \citenamefont {Lee}(2019)}]{LeeJY2019-2}%
    \BibitemOpen
    \bibfield  {author} {\bibinfo {author} {\bibfnamefont {H.}~\bibnamefont
    {Zhou}}\ and\ \bibinfo {author} {\bibfnamefont {J.~Y.}\ \bibnamefont {Lee}},\
    }\bibfield  {title} {\bibinfo {title} {Periodic table for topological bands
    with non-Hermitian symmetries},\ }\href {\doibase/10.1103/PhysRevB.99.235112}
    {\bibfield  {journal} {\bibinfo  {journal} {Phys. Rev. B}\ }\textbf {\bibinfo
    {volume} {99}},\ \bibinfo {pages} {235112} (\bibinfo {year}
    {2019})}\BibitemShut {NoStop}%
  \bibitem [{\citenamefont {Xiao}\ \emph {et~al.}(2020)\citenamefont {Xiao},
    \citenamefont {Deng}, \citenamefont {Wang}, \citenamefont {Zhu},
    \citenamefont {Wang}, \citenamefont {Yi},\ and\ \citenamefont
    {Xue}}]{Xiao2020}%
    \BibitemOpen
    \bibfield  {author} {\bibinfo {author} {\bibfnamefont {L.}~\bibnamefont
    {Xiao}}, \bibinfo {author} {\bibfnamefont {T.}~\bibnamefont {Deng}}, \bibinfo
    {author} {\bibfnamefont {K.}~\bibnamefont {Wang}}, \bibinfo {author}
    {\bibfnamefont {G.}~\bibnamefont {Zhu}}, \bibinfo {author} {\bibfnamefont
    {Z.}~\bibnamefont {Wang}}, \bibinfo {author} {\bibfnamefont {W.}~\bibnamefont
    {Yi}}, \ and\ \bibinfo {author} {\bibfnamefont {P.}~\bibnamefont {Xue}},\
    }\bibfield  {title} {\bibinfo {title} {Non-Hermitian bulk--boundary
    correspondence in quantum dynamics},\ }\href
    {\doibase/10.1038/s41567-020-0836-6} {\bibfield  {journal} {\bibinfo
    {journal} {Nature Physics}\ }\textbf {\bibinfo {volume} {16}},\ \bibinfo
    {pages} {761} (\bibinfo {year} {2020})}\BibitemShut {NoStop}%
  \bibitem [{\citenamefont {Brody}(2013)}]{Brody2013}%
    \BibitemOpen
    \bibfield  {author} {\bibinfo {author} {\bibfnamefont {D.~C.}\ \bibnamefont
    {Brody}},\ }\bibfield  {title} {\bibinfo {title} {Biorthogonal quantum
    mechanics},\ }\href {\doibase/10.1088/1751-8113/47/3/035305} {\bibfield
    {journal} {\bibinfo  {journal} {Journal of Physics A: Mathematical and
    Theoretical}\ }\textbf {\bibinfo {volume} {47}},\ \bibinfo {pages} {035305}
    (\bibinfo {year} {2013})}\BibitemShut {NoStop}%
  \bibitem [{\citenamefont {Scholtz}\ \emph {et~al.}(1992)\citenamefont
    {Scholtz}, \citenamefont {Geyer},\ and\ \citenamefont {Hahne}}]{Scholtz1992}%
    \BibitemOpen
    \bibfield  {author} {\bibinfo {author} {\bibfnamefont {F.}~\bibnamefont
    {Scholtz}}, \bibinfo {author} {\bibfnamefont {H.}~\bibnamefont {Geyer}}, \
    and\ \bibinfo {author} {\bibfnamefont {F.}~\bibnamefont {Hahne}},\ }\bibfield
     {title} {\bibinfo {title} {Quasi-Hermitian operators in quantum mechanics
    and the variational principle},\ }\href
    {\doibase/10.1016/0003-4916(92)90284-s} {\bibfield  {journal} {\bibinfo
    {journal} {Annals of Physics}\ }\textbf {\bibinfo {volume} {213}},\ \bibinfo
    {pages} {74} (\bibinfo {year} {1992})}\BibitemShut {NoStop}%
  \bibitem [{\citenamefont {Mostafazadeh}(2010)}]{Mostafazadeh2010}%
    \BibitemOpen
    \bibfield  {author} {\bibinfo {author} {\bibfnamefont {A.}~\bibnamefont
    {Mostafazadeh}},\ }\bibfield  {title} {\bibinfo {title} {{Pseudo}-{Hermitian}
    {Representation} {of} {Quantum} {Mechanics}},\ }\href
    {\doibase/10.1142/s0219887810004816} {\bibfield  {journal} {\bibinfo
    {journal} {International Journal of Geometric Methods in Modern Physics}\
    }\textbf {\bibinfo {volume} {07}},\ \bibinfo {pages} {1191} (\bibinfo {year}
    {2010})}\BibitemShut {NoStop}%
  \bibitem [{\citenamefont {Dorey}\ \emph {et~al.}(2004)\citenamefont {Dorey},
    \citenamefont {Dunning},\ and\ \citenamefont {Tateo}}]{Dorey2004}%
    \BibitemOpen
    \bibfield  {author} {\bibinfo {author} {\bibfnamefont {P.}~\bibnamefont
    {Dorey}}, \bibinfo {author} {\bibfnamefont {C.}~\bibnamefont {Dunning}}, \
    and\ \bibinfo {author} {\bibfnamefont {R.}~\bibnamefont {Tateo}},\ }\bibfield
     {title} {\bibinfo {title} {A reality proof in {PT}-symmetric quantum
    mechanics},\ }\href {\doibase/10.1023/b:cjop.0000014365.19507.b6} {\bibfield
    {journal} {\bibinfo  {journal} {Czechoslovak Journal of Physics}\ }\textbf
    {\bibinfo {volume} {54}},\ \bibinfo {pages} {35} (\bibinfo {year}
    {2004})}\BibitemShut {NoStop}%
  \bibitem [{\citenamefont {Wen}\ and\ \citenamefont {Zee}(1992)}]{Wen1992}%
    \BibitemOpen
    \bibfield  {author} {\bibinfo {author} {\bibfnamefont {X.~G.}\ \bibnamefont
    {Wen}}\ and\ \bibinfo {author} {\bibfnamefont {A.}~\bibnamefont {Zee}},\
    }\bibfield  {title} {\bibinfo {title} {Shift and spin vector: New topological
    quantum numbers for the Hall fluids},\ }\href
    {\doibase/10.1103/PhysRevLett.69.953} {\bibfield  {journal} {\bibinfo
    {journal} {Phys. Rev. Lett.}\ }\textbf {\bibinfo {volume} {69}},\ \bibinfo
    {pages} {953} (\bibinfo {year} {1992})}\BibitemShut {NoStop}%
  \bibitem [{\citenamefont {Avron}\ \emph {et~al.}(1995)\citenamefont {Avron},
    \citenamefont {Seiler},\ and\ \citenamefont {Zograf}}]{Zograf1995}%
    \BibitemOpen
    \bibfield  {author} {\bibinfo {author} {\bibfnamefont {J.~E.}\ \bibnamefont
    {Avron}}, \bibinfo {author} {\bibfnamefont {R.}~\bibnamefont {Seiler}}, \
    and\ \bibinfo {author} {\bibfnamefont {P.~G.}\ \bibnamefont {Zograf}},\
    }\bibfield  {title} {\bibinfo {title} {Viscosity of quantum Hall fluids},\
    }\href {\doibase/10.1103/PhysRevLett.75.697} {\bibfield  {journal} {\bibinfo
    {journal} {Phys. Rev. Lett.}\ }\textbf {\bibinfo {volume} {75}},\ \bibinfo
    {pages} {697} (\bibinfo {year} {1995})}\BibitemShut {NoStop}%
  \bibitem [{\citenamefont {Can}\ \emph {et~al.}(2014)\citenamefont {Can},
    \citenamefont {Laskin},\ and\ \citenamefont {Wiegmann}}]{Wiegmann2014}%
    \BibitemOpen
    \bibfield  {author} {\bibinfo {author} {\bibfnamefont {T.}~\bibnamefont
    {Can}}, \bibinfo {author} {\bibfnamefont {M.}~\bibnamefont {Laskin}}, \ and\
    \bibinfo {author} {\bibfnamefont {P.}~\bibnamefont {Wiegmann}},\ }\bibfield
    {title} {\bibinfo {title} {Fractional quantum Hall effect in a curved space:
    Gravitational anomaly and electromagnetic response},\ }\href
    {\doibase/10.1103/PhysRevLett.113.046803} {\bibfield  {journal} {\bibinfo
    {journal} {Phys. Rev. Lett.}\ }\textbf {\bibinfo {volume} {113}},\ \bibinfo
    {pages} {046803} (\bibinfo {year} {2014})}\BibitemShut {NoStop}%
  \bibitem [{\citenamefont {Schine}\ \emph {et~al.}(2019)\citenamefont {Schine},
    \citenamefont {Chalupnik}, \citenamefont {Can}, \citenamefont {Gromov},\ and\
    \citenamefont {Simon}}]{Schine2019}%
    \BibitemOpen
    \bibfield  {author} {\bibinfo {author} {\bibfnamefont {N.}~\bibnamefont
    {Schine}}, \bibinfo {author} {\bibfnamefont {M.}~\bibnamefont {Chalupnik}},
    \bibinfo {author} {\bibfnamefont {T.}~\bibnamefont {Can}}, \bibinfo {author}
    {\bibfnamefont {A.}~\bibnamefont {Gromov}}, \ and\ \bibinfo {author}
    {\bibfnamefont {J.}~\bibnamefont {Simon}},\ }\bibfield  {title} {\bibinfo
    {title} {Electromagnetic and gravitational responses of photonic Landau
    levels},\ }\href {\doibase/10.1038/s41586-018-0817-4} {\bibfield  {journal}
    {\bibinfo  {journal} {Nature}\ }\textbf {\bibinfo {volume} {565}},\ \bibinfo
    {pages} {173} (\bibinfo {year} {2019})}\BibitemShut {NoStop}%
  \bibitem [{\citenamefont {Bekenstein}\ \emph {et~al.}(2017)\citenamefont
    {Bekenstein}, \citenamefont {Kabessa}, \citenamefont {Sharabi}, \citenamefont
    {Tal}, \citenamefont {Engheta}, \citenamefont {Eisenstein}, \citenamefont
    {Agranat},\ and\ \citenamefont {Segev}}]{Bekenstein2017}%
    \BibitemOpen
    \bibfield  {author} {\bibinfo {author} {\bibfnamefont {R.}~\bibnamefont
    {Bekenstein}}, \bibinfo {author} {\bibfnamefont {Y.}~\bibnamefont {Kabessa}},
    \bibinfo {author} {\bibfnamefont {Y.}~\bibnamefont {Sharabi}}, \bibinfo
    {author} {\bibfnamefont {O.}~\bibnamefont {Tal}}, \bibinfo {author}
    {\bibfnamefont {N.}~\bibnamefont {Engheta}}, \bibinfo {author} {\bibfnamefont
    {G.}~\bibnamefont {Eisenstein}}, \bibinfo {author} {\bibfnamefont {A.~J.}\
    \bibnamefont {Agranat}}, \ and\ \bibinfo {author} {\bibfnamefont
    {M.}~\bibnamefont {Segev}},\ }\bibfield  {title} {\bibinfo {title} {Control
    of light by curved space in nanophotonic structures},\ }\href
    {\doibase/10.1038/s41566-017-0008-0} {\bibfield  {journal} {\bibinfo
    {journal} {Nature Photonics}\ }\textbf {\bibinfo {volume} {11}},\ \bibinfo
    {pages} {664} (\bibinfo {year} {2017})}\BibitemShut {NoStop}%
  \bibitem [{\citenamefont {Zhou}\ \emph {et~al.}(2018)\citenamefont {Zhou},
    \citenamefont {Wu}, \citenamefont {Guo}, \citenamefont {Wang}, \citenamefont
    {Pu},\ and\ \citenamefont {Zhou}}]{ZhouZW2018}%
    \BibitemOpen
    \bibfield  {author} {\bibinfo {author} {\bibfnamefont {X.-F.}\ \bibnamefont
    {Zhou}}, \bibinfo {author} {\bibfnamefont {C.}~\bibnamefont {Wu}}, \bibinfo
    {author} {\bibfnamefont {G.-C.}\ \bibnamefont {Guo}}, \bibinfo {author}
    {\bibfnamefont {R.}~\bibnamefont {Wang}}, \bibinfo {author} {\bibfnamefont
    {H.}~\bibnamefont {Pu}}, \ and\ \bibinfo {author} {\bibfnamefont {Z.-W.}\
    \bibnamefont {Zhou}},\ }\bibfield  {title} {\bibinfo {title} {Synthetic
    Landau levels and spinor vortex matter on a Haldane spherical surface with a
    magnetic monopole},\ }\href {\doibase/10.1103/PhysRevLett.120.130402}
    {\bibfield  {journal} {\bibinfo  {journal} {Phys. Rev. Lett.}\ }\textbf
    {\bibinfo {volume} {120}},\ \bibinfo {pages} {130402} (\bibinfo {year}
    {2018})}\BibitemShut {NoStop}%
  \bibitem [{\citenamefont {Koll{\'{a}}r}\ \emph {et~al.}(2019)\citenamefont
    {Koll{\'{a}}r}, \citenamefont {Fitzpatrick},\ and\ \citenamefont
    {Houck}}]{Kollr2019}%
    \BibitemOpen
    \bibfield  {author} {\bibinfo {author} {\bibfnamefont {A.~J.}\ \bibnamefont
    {Koll{\'{a}}r}}, \bibinfo {author} {\bibfnamefont {M.}~\bibnamefont
    {Fitzpatrick}}, \ and\ \bibinfo {author} {\bibfnamefont {A.~A.}\ \bibnamefont
    {Houck}},\ }\bibfield  {title} {\bibinfo {title} {Hyperbolic lattices in
    circuit quantum electrodynamics},\ }\href
    {\doibase/10.1038/s41586-019-1348-3} {\bibfield  {journal} {\bibinfo
    {journal} {Nature}\ }\textbf {\bibinfo {volume} {571}},\ \bibinfo {pages}
    {45} (\bibinfo {year} {2019})}\BibitemShut {NoStop}%
  \bibitem [{\citenamefont {Hatano}\ and\ \citenamefont
    {Nelson}(1996)}]{Hatano1996}%
    \BibitemOpen
    \bibfield  {author} {\bibinfo {author} {\bibfnamefont {N.}~\bibnamefont
    {Hatano}}\ and\ \bibinfo {author} {\bibfnamefont {D.~R.}\ \bibnamefont
    {Nelson}},\ }\bibfield  {title} {\bibinfo {title} {Localization transitions
    in non-Hermitian quantum mechanics},\ }\href
    {\doibase/10.1103/PhysRevLett.77.570} {\bibfield  {journal} {\bibinfo
    {journal} {Phys. Rev. Lett.}\ }\textbf {\bibinfo {volume} {77}},\ \bibinfo
    {pages} {570} (\bibinfo {year} {1996})}\BibitemShut {NoStop}%
  \bibitem [{\citenamefont {Nelson}\ and\ \citenamefont
    {Vinokur}(1993)}]{Nelson1993}%
    \BibitemOpen
    \bibfield  {author} {\bibinfo {author} {\bibfnamefont {D.~R.}\ \bibnamefont
    {Nelson}}\ and\ \bibinfo {author} {\bibfnamefont {V.~M.}\ \bibnamefont
    {Vinokur}},\ }\bibfield  {title} {\bibinfo {title} {Boson localization and
    correlated pinning of superconducting vortex arrays},\ }\href
    {\doibase/10.1103/PhysRevB.48.13060} {\bibfield  {journal} {\bibinfo
    {journal} {Phys. Rev. B}\ }\textbf {\bibinfo {volume} {48}},\ \bibinfo
    {pages} {13060} (\bibinfo {year} {1993})}\BibitemShut {NoStop}%
  \bibitem [{\citenamefont {Amir}\ \emph {et~al.}(2016)\citenamefont {Amir},
    \citenamefont {Hatano},\ and\ \citenamefont {Nelson}}]{Nelson2016}%
    \BibitemOpen
    \bibfield  {author} {\bibinfo {author} {\bibfnamefont {A.}~\bibnamefont
    {Amir}}, \bibinfo {author} {\bibfnamefont {N.}~\bibnamefont {Hatano}}, \ and\
    \bibinfo {author} {\bibfnamefont {D.~R.}\ \bibnamefont {Nelson}},\ }\bibfield
     {title} {\bibinfo {title} {Non-Hermitian localization in biological
    networks},\ }\href {\doibase/10.1103/PhysRevE.93.042310} {\bibfield
    {journal} {\bibinfo  {journal} {Phys. Rev. E}\ }\textbf {\bibinfo {volume}
    {93}},\ \bibinfo {pages} {042310} (\bibinfo {year} {2016})}\BibitemShut
    {NoStop}%
  \bibitem [{\citenamefont {Lodahl}\ \emph {et~al.}(2017)\citenamefont {Lodahl},
    \citenamefont {Mahmoodian}, \citenamefont {Stobbe}, \citenamefont
    {Rauschenbeutel}, \citenamefont {Schneeweiss}, \citenamefont {Volz},
    \citenamefont {Pichler},\ and\ \citenamefont {Zoller}}]{Zoller2017}%
    \BibitemOpen
    \bibfield  {author} {\bibinfo {author} {\bibfnamefont {P.}~\bibnamefont
    {Lodahl}}, \bibinfo {author} {\bibfnamefont {S.}~\bibnamefont {Mahmoodian}},
    \bibinfo {author} {\bibfnamefont {S.}~\bibnamefont {Stobbe}}, \bibinfo
    {author} {\bibfnamefont {A.}~\bibnamefont {Rauschenbeutel}}, \bibinfo
    {author} {\bibfnamefont {P.}~\bibnamefont {Schneeweiss}}, \bibinfo {author}
    {\bibfnamefont {J.}~\bibnamefont {Volz}}, \bibinfo {author} {\bibfnamefont
    {H.}~\bibnamefont {Pichler}}, \ and\ \bibinfo {author} {\bibfnamefont
    {P.}~\bibnamefont {Zoller}},\ }\bibfield  {title} {\bibinfo {title} {Chiral
    quantum optics},\ }\href {\doibase/10.1038/nature21037} {\bibfield  {journal}
    {\bibinfo  {journal} {Nature}\ }\textbf {\bibinfo {volume} {541}},\ \bibinfo
    {pages} {473} (\bibinfo {year} {2017})}\BibitemShut {NoStop}%
  \bibitem [{\citenamefont {Yang}\ \emph {et~al.}(2021)\citenamefont {Yang},
    \citenamefont {Naaman}, \citenamefont {Paltiel},\ and\ \citenamefont
    {Parkin}}]{Yang2021}%
    \BibitemOpen
    \bibfield  {author} {\bibinfo {author} {\bibfnamefont {S.-H.}\ \bibnamefont
    {Yang}}, \bibinfo {author} {\bibfnamefont {R.}~\bibnamefont {Naaman}},
    \bibinfo {author} {\bibfnamefont {Y.}~\bibnamefont {Paltiel}}, \ and\
    \bibinfo {author} {\bibfnamefont {S.~S.~P.}\ \bibnamefont {Parkin}},\
    }\bibfield  {title} {\bibinfo {title} {Chiral spintronics},\ }\href
    {\doibase/10.1038/s42254-021-00302-9} {\bibfield  {journal} {\bibinfo
    {journal} {Nature Reviews Physics}\ }\textbf {\bibinfo {volume} {3}},\
    \bibinfo {pages} {328} (\bibinfo {year} {2021})}\BibitemShut {NoStop}%
  \bibitem [{\citenamefont {Carroll}(2019)}]{Carroll2019}%
    \BibitemOpen
    \bibfield  {author} {\bibinfo {author} {\bibfnamefont {S.~M.}\ \bibnamefont
    {Carroll}},\ }\href {\doibase/10.1017/9781108770385} {\emph {\bibinfo {title}
    {Spacetime and Geometry}}}\ (\bibinfo  {publisher} {Cambridge University
    Press},\ \bibinfo {year} {2019})\BibitemShut {NoStop}%
  \bibitem [{\citenamefont {Bender}\ \emph {et~al.}(2005)\citenamefont {Bender},
    \citenamefont {Brandt}, \citenamefont {Chen},\ and\ \citenamefont
    {Wang}}]{Bender2005}%
    \BibitemOpen
    \bibfield  {author} {\bibinfo {author} {\bibfnamefont {C.~M.}\ \bibnamefont
    {Bender}}, \bibinfo {author} {\bibfnamefont {S.~F.}\ \bibnamefont {Brandt}},
    \bibinfo {author} {\bibfnamefont {J.-H.}\ \bibnamefont {Chen}}, \ and\
    \bibinfo {author} {\bibfnamefont {Q.}~\bibnamefont {Wang}},\ }\bibfield
    {title} {\bibinfo {title} {Ghost busting: $\mathcal{P}\mathcal{T}$-symmetric
    interpretation of the Lee model},\ }\href
    {\doibase/10.1103/PhysRevD.71.025014} {\bibfield  {journal} {\bibinfo
    {journal} {Phys. Rev. D}\ }\textbf {\bibinfo {volume} {71}},\ \bibinfo
    {pages} {025014} (\bibinfo {year} {2005})}\BibitemShut {NoStop}%
  \bibitem [{\citenamefont {Zhang}\ \emph {et~al.}(2021)\citenamefont {Zhang},
    \citenamefont {Lv}, \citenamefont {Yan},\ and\ \citenamefont
    {Zhou}}]{Zhou2020}%
    \BibitemOpen
    \bibfield  {author} {\bibinfo {author} {\bibfnamefont {R.}~\bibnamefont
    {Zhang}}, \bibinfo {author} {\bibfnamefont {C.}~\bibnamefont {Lv}}, \bibinfo
    {author} {\bibfnamefont {Y.}~\bibnamefont {Yan}}, \ and\ \bibinfo {author}
    {\bibfnamefont {Q.}~\bibnamefont {Zhou}},\ }\bibfield  {title} {\bibinfo
    {title} {Efimov-like states and quantum funneling effects on synthetic
    hyperbolic surfaces},\ }\href {\doibase/10.1016/j.scib.2021.06.017}
    {\bibfield  {journal} {\bibinfo  {journal} {Science Bulletin}\ } (\bibinfo
    {year} {2021}),\ 10.1016/j.scib.2021.06.017}\BibitemShut {NoStop}%
  \bibitem [{\citenamefont {Yokomizo}\ and\ \citenamefont
    {Murakami}(2019)}]{Yokomizo2019}%
    \BibitemOpen
    \bibfield  {author} {\bibinfo {author} {\bibfnamefont {K.}~\bibnamefont
    {Yokomizo}}\ and\ \bibinfo {author} {\bibfnamefont {S.}~\bibnamefont
    {Murakami}},\ }\bibfield  {title} {\bibinfo {title} {Non-Bloch band theory of
    non-Hermitian systems},\ }\href {\doibase/10.1103/PhysRevLett.123.066404}
    {\bibfield  {journal} {\bibinfo  {journal} {Phys. Rev. Lett.}\ }\textbf
    {\bibinfo {volume} {123}},\ \bibinfo {pages} {066404} (\bibinfo {year}
    {2019})}\BibitemShut {NoStop}%
  \bibitem [{\citenamefont {Yang}\ \emph {et~al.}(2020)\citenamefont {Yang},
    \citenamefont {Zhang}, \citenamefont {Fang},\ and\ \citenamefont
    {Hu}}]{Hu2020}%
    \BibitemOpen
    \bibfield  {author} {\bibinfo {author} {\bibfnamefont {Z.}~\bibnamefont
    {Yang}}, \bibinfo {author} {\bibfnamefont {K.}~\bibnamefont {Zhang}},
    \bibinfo {author} {\bibfnamefont {C.}~\bibnamefont {Fang}}, \ and\ \bibinfo
    {author} {\bibfnamefont {J.}~\bibnamefont {Hu}},\ }\bibfield  {title}
    {\bibinfo {title} {Non-Hermitian bulk-boundary correspondence and auxiliary
    generalized Brillouin zone theory},\ }\href
    {\doibase/10.1103/PhysRevLett.125.226402} {\bibfield  {journal} {\bibinfo
    {journal} {Phys. Rev. Lett.}\ }\textbf {\bibinfo {volume} {125}},\ \bibinfo
    {pages} {226402} (\bibinfo {year} {2020})}\BibitemShut {NoStop}%
  \bibitem [{\citenamefont {Comtet}(1987)}]{Comtet1987}%
    \BibitemOpen
    \bibfield  {author} {\bibinfo {author} {\bibfnamefont {A.}~\bibnamefont
    {Comtet}},\ }\bibfield  {title} {\bibinfo {title} {On the Landau levels on
    the hyperbolic plane},\ }\href {\doibase/10.1016/0003-4916(87)90098-4}
    {\bibfield  {journal} {\bibinfo  {journal} {Annals of Physics}\ }\textbf
    {\bibinfo {volume} {173}},\ \bibinfo {pages} {185} (\bibinfo {year}
    {1987})}\BibitemShut {NoStop}%
  \bibitem [{\citenamefont {Hofstadter}(1976)}]{Hofstadter1976}%
    \BibitemOpen
    \bibfield  {author} {\bibinfo {author} {\bibfnamefont {D.~R.}\ \bibnamefont
    {Hofstadter}},\ }\bibfield  {title} {\bibinfo {title} {Energy levels and wave
    functions of Bloch electrons in rational and irrational magnetic fields},\
    }\href {\doibase/10.1103/PhysRevB.14.2239} {\bibfield  {journal} {\bibinfo
    {journal} {Phys. Rev. B}\ }\textbf {\bibinfo {volume} {14}},\ \bibinfo
    {pages} {2239} (\bibinfo {year} {1976})}\BibitemShut {NoStop}%
  \end{thebibliography}

\begin{thebibliography}{7}%
   \makeatletter
   \providecommand \@ifxundefined [1]{%
    \@ifx{#1\undefined}
   }%
   \providecommand \@ifnum [1]{%
    \ifnum #1\expandafter \@firstoftwo
    \else \expandafter \@secondoftwo
    \fi
   }%
   \providecommand \@ifx [1]{%
    \ifx #1\expandafter \@firstoftwo
    \else \expandafter \@secondoftwo
    \fi
   }%
   \providecommand \natexlab [1]{#1}%
   \providecommand \enquote  [1]{#1}%
   \providecommand \bibnamefont  [1]{#1}%
   \providecommand \bibfnamefont [1]{#1}%
   \providecommand \citenamefont [1]{#1}%
   \providecommand \href@noop [0]{\@secondoftwo}%
   \providecommand \href [0]{\begingroup \@sanitize@url \@href}%
   \providecommand \@href[1]{\@@startlink{#1}\@@href}%
   \providecommand \@@href[1]{\endgroup#1\@@endlink}%
   \providecommand \@sanitize@url [0]{\catcode `\\12\catcode `\$12\catcode
     `\&12\catcode `\#12\catcode `\^12\catcode `\_12\catcode `\%12\relax}%
   \providecommand \@@startlink[1]{}%
   \providecommand \@@endlink[0]{}%
   \providecommand \url  [0]{\begingroup\@sanitize@url \@url }%
   \providecommand \@url [1]{\endgroup\@href {#1}{\urlprefix }}%
   \providecommand \urlprefix  [0]{URL }%
   \providecommand \Eprint [0]{\href }%
   \providecommand \doibase [0]{https://dx.doi.org}%
   \providecommand \selectlanguage [0]{\@gobble}%
   \providecommand \bibinfo  [0]{\@secondoftwo}%
   \providecommand \bibfield  [0]{\@secondoftwo}%
   \providecommand \translation [1]{[#1]}%
   \providecommand \BibitemOpen [0]{}%
   \providecommand \bibitemStop [0]{}%
   \providecommand \bibitemNoStop [0]{.\EOS\space}%
   \providecommand \EOS [0]{\spacefactor3000\relax}%
   \providecommand \BibitemShut  [1]{\csname bibitem#1\endcsname}%
   \let\auto@bib@innerbib\@empty
   \bibitem [{\citenamefont {Carroll}(2019)}]{SCarroll2019}%
     \BibitemOpen
     \bibfield  {author} {\bibinfo {author} {\bibfnamefont {S.~M.}\ \bibnamefont
     {Carroll}},\ }\href@noop {} {\emph {\bibinfo {title} {Spacetime and
     Geometry}}}\ (\bibinfo  {publisher} {Cambridge University Press},\ \bibinfo
     {year} {2019})\BibitemShut {NoStop}%
   \bibitem [{\citenamefont {Gutzwiller}(1983)}]{SGutzwiller1983}%
     \BibitemOpen
     \bibfield  {author} {\bibinfo {author} {\bibfnamefont {M.~C.}\ \bibnamefont
     {Gutzwiller}},\ }\bibfield  {title} {\bibinfo {title} {Stochastic behavior in
     quantum scattering},\ }\href {\doibase/10.1016/0167-2789(83)90138-0}
     {\bibfield  {journal} {\bibinfo  {journal} {Physica D: Nonlinear Phenomena}\
     }\textbf {\bibinfo {volume} {7}},\ \bibinfo {pages} {341} (\bibinfo {year}
     {1983})}\BibitemShut {NoStop}%
   \bibitem [{\citenamefont {da~Costa}(1981)}]{SCosta1982}%
     \BibitemOpen
     \bibfield  {author} {\bibinfo {author} {\bibfnamefont {R.~C.~T.}\
     \bibnamefont {da~Costa}},\ }\bibfield  {title} {\bibinfo {title} {Quantum
     mechanics of a constrained particle},\ }\href
     {\doibase/10.1103/PhysRevA.23.1982} {\bibfield  {journal} {\bibinfo
     {journal} {Phys. Rev. A}\ }\textbf {\bibinfo {volume} {23}},\ \bibinfo
     {pages} {1982} (\bibinfo {year} {1981})}\BibitemShut {NoStop}%
   \bibitem [{\citenamefont {Yao}\ and\ \citenamefont {Wang}(2018)}]{SWang2018}%
     \BibitemOpen
     \bibfield  {author} {\bibinfo {author} {\bibfnamefont {S.}~\bibnamefont
     {Yao}}\ and\ \bibinfo {author} {\bibfnamefont {Z.}~\bibnamefont {Wang}},\
     }\bibfield  {title} {\bibinfo {title} {Edge states and topological invariants
     of non-Hermitian systems},\ }\href {\doibase/10.1103/PhysRevLett.121.086803}
     {\bibfield  {journal} {\bibinfo  {journal} {Phys. Rev. Lett.}\ }\textbf
     {\bibinfo {volume} {121}},\ \bibinfo {pages} {086803} (\bibinfo {year}
     {2018})}\BibitemShut {NoStop}%
   \bibitem [{\citenamefont {Yokomizo}\ and\ \citenamefont
     {Murakami}(2019)}]{SYokomizo2019}%
     \BibitemOpen
     \bibfield  {author} {\bibinfo {author} {\bibfnamefont {K.}~\bibnamefont
     {Yokomizo}}\ and\ \bibinfo {author} {\bibfnamefont {S.}~\bibnamefont
     {Murakami}},\ }\bibfield  {title} {\bibinfo {title} {Non-Bloch band theory of
     non-Hermitian systems},\ }\href {\doibase/10.1103/PhysRevLett.123.066404}
     {\bibfield  {journal} {\bibinfo  {journal} {Phys. Rev. Lett.}\ }\textbf
     {\bibinfo {volume} {123}},\ \bibinfo {pages} {066404} (\bibinfo {year}
     {2019})}\BibitemShut {NoStop}%
   \bibitem [{\citenamefont {Yang}\ \emph {et~al.}(2020)\citenamefont {Yang},
     \citenamefont {Zhang}, \citenamefont {Fang},\ and\ \citenamefont
     {Hu}}]{SHu2020}%
     \BibitemOpen
     \bibfield  {author} {\bibinfo {author} {\bibfnamefont {Z.}~\bibnamefont
     {Yang}}, \bibinfo {author} {\bibfnamefont {K.}~\bibnamefont {Zhang}},
     \bibinfo {author} {\bibfnamefont {C.}~\bibnamefont {Fang}}, \ and\ \bibinfo
     {author} {\bibfnamefont {J.}~\bibnamefont {Hu}},\ }\bibfield  {title}
     {\bibinfo {title} {Non-Hermitian bulk-boundary correspondence and auxiliary
     generalized Brillouin zone theory},\ }\href
     {\doibase/10.1103/PhysRevLett.125.226402} {\bibfield  {journal} {\bibinfo
     {journal} {Phys. Rev. Lett.}\ }\textbf {\bibinfo {volume} {125}},\ \bibinfo
     {pages} {226402} (\bibinfo {year} {2020})}\BibitemShut {NoStop}%
   \bibitem [{\citenamefont {Hofstadter}(1976)}]{SHofstadter1976}%
     \BibitemOpen
     \bibfield  {author} {\bibinfo {author} {\bibfnamefont {D.~R.}\ \bibnamefont
     {Hofstadter}},\ }\bibfield  {title} {\bibinfo {title} {Energy levels and wave
     functions of Bloch electrons in rational and irrational magnetic fields},\
     }\href {\doibase/10.1103/PhysRevB.14.2239} {\bibfield  {journal} {\bibinfo
     {journal} {Phys. Rev. B}\ }\textbf {\bibinfo {volume} {14}},\ \bibinfo
     {pages} {2239} (\bibinfo {year} {1976})}\BibitemShut {NoStop}%
   \end{thebibliography}
%

\end{document}